\documentclass[aps,prd,showpacs,preprint,showpacs,showkeys,preprintnumbers,superscriptaddress,amsmath,amssymb,nofootinbib]{revtex4-1}
\usepackage{amsmath}
\usepackage{latexsym}
\usepackage{pstricks,pst-coil}
\usepackage{pst-all}
\def\uma{\rm 1\!\!\hskip 1 pt l}
\usepackage{color}
\usepackage{latexsym}
\usepackage{amsmath}
\usepackage{amssymb}
\usepackage{euscript}
\usepackage{pstricks}
\usepackage{graphics}
\usepackage{graphicx}
\usepackage{picture}

\newcommand{\be}{\begin{equation}}
\newcommand{\ee}{\end{equation}}
\newcommand{\ba}{\begin{eqnarray}}
\newcommand{\ea}{\end{eqnarray}}
\newcommand{\p}{\partial}
\def\ni{\noindent}

\begin{document}

\title{\Large The Yang-Mills gauge theory in DFR noncommutative space-time}

\author{Everton M. C. Abreu}\email{evertonabreu@ufrrj.br}
\affiliation{Grupo de F\' isica Te\'orica e Matem\'atica F\' isica, Departamento de F\'{i}sica, Universidade Federal Rural do Rio de Janeiro, 23890-971, Serop\'edica - RJ, Brasil}
\affiliation{Departamento de F\'{i}sica, Universidade Federal de Juiz de Fora, 36036-330, Juiz de Fora - MG, Brasil}
\author{M. J. Neves}\email{mariojr@ufrrj.br}
\affiliation{Grupo de F\' isica Te\'orica e Matem\'atica F\' isica, Departamento de F\'{i}sica, Universidade Federal Rural do Rio de Janeiro, 23890-971, Serop\'edica - RJ, Brasil}

\date{\today}

\begin{abstract}
\ni The Doplicher-Fredenhagen-Roberts (DFR) framework for noncommutative (NC) space-times
is considered as an alternative approach to describe the physics of quantum gravity, for instance. In this formalism,
the NC parameter, {\it i.e.} $\theta^{\mu\nu}$, is promoted to a coordinate of a new extended space-time.
Consequently, we have a field theory in a space-time with spatial extra-dimensions. This new coordinate
has a canonical momentum associated, where the effects of a new physics can emerge in the fields
propagation along the extra-dimension. In this paper we introduce the gauge invariance in the DFR NC
space-time. We present the non-Abelian gauge symmetry in DFR formalism, and the consequences of this symmetry
in the presence of such extra-dimension.
The gauge symmetry in this DFR scenario can reveal new fields attached to $\theta$-extra-dimension.
We obtain the propagation of these gauge fields in terms of canonical momentum
associated with $\theta$-coordinate.
\end{abstract}



\maketitle

\pagestyle{myheadings}
\markright{Yang-Mills gauge theory in DFR noncommutative space-time}


\section{Introduction}

The inconvenience of having infinities that destroy the final results of several calculations in QFT have motivated theoretical physicists to ask if a continuum space-time would be really necessary. The alternative would be to construct a discrete space-time with a noncommutative (NC) algebra,
where the position coordinates are proportional to operators $\hat{X}^{\mu}\,(\mu=0,1,2,3)$ and they must satisfy the commutation relations
\begin{eqnarray}\label{xmuxnu}
\left[\,\hat{X}^\mu\,,\,\hat{X}^\nu\,\right]\,=\,i\,\ell \, \theta^{\mu\nu} \, \hat{\uma} \,\,,
\end{eqnarray}
where $\ell$ is a length parameter, $\theta^{\mu\nu}$ is an anti-symmetric constant matrix and $\hat{\uma}$, the identity operator. Putting these ideas all together, Snyder \cite{snyder47} published the first work considering the space-time as being a NC one. However, Yang \cite{yang47} demonstrated that Snyder's hopes about the disappearance of the infinities were not achieved by noncommutativity.  This fact has doomed Snyder NC theory to years of ostracism.  After the string theory important result that the algebra obtained from the case of a string theory embedded in a magnetic background is NC, a new wave concerning noncommutativity was rekindle \cite{seibergwitten99}. In current days, the NC approach is also a subject discussed at the quantum gravity level \cite{QG4,QG5}.

One of the paths of introducing noncommutativity is through the Moyal-Weyl (or star) product where the NC parameter,
{\it i.e.} $\theta^{\mu\nu}$, is an anti-symmetric constant. However, at superior orders of calculations, the Moyal-Weyl product turns out to be
highly nonlocal. This fact has lead us to work with low orders in $\theta^{\mu\nu}$. Although it maintains the translational invariance,
the Lorentz symmetry is not preserved \cite{Szabo03}. For example, in the case of the hydrogen atom, it breaks the rotational symmetry of the model, which removes the degeneracy of the energy levels \cite{Chaichian}.

One way to heal this problem was introduced by Doplicher, Fredenhagen and Roberts (DFR) which have promoted the parameter $\theta^{\mu\nu}$ to the role of an ordinary coordinate of the system \cite{DFR1,DFR2}. This so-called extended and new NC space-time has ten dimensions: four
relative to Minkowski space-time ordinary positions and six relative to $\theta$-space. Recently, in \cite{Morita} the authors have conjectured to construct a DFR  space-time extension, introducing the conjugate canonical momentum associated with $\theta^{\mu\nu}$ \cite{Amorim1} (for a review, the reader can see \cite{amo}). This new framework would be characterized by a field theory constructed in a space-time with extra-dimensions $(4+6)$.   Besides, it would
not need necessarily the presence of a length scale $\ell$ localized into the six dimensions
of the $\theta$-space where, from (\ref{xmuxnu}) we can see that $\theta^{\mu\nu}$ would have dimension of length-square (a kind of Planck area),
when we make $\ell=1$. The length scale can be introduced directly in the algebra, and taking the limit with no such scale, the usual algebra of
the commutative space-time is recovered. Besides the Lorentz invariance was also recovered, and obviously we hope that causality aspects in
QFT in this $\left(x+\theta\right)$ space-time must be preserved too \cite{EMCAbreuMJNeves2012}.

In this approach, as we have said, the parameter $\theta^{\mu\nu}$ is also promoted to a position operator, say $\hat{\theta}^{\mu\nu}$, that participate of the algebra, and it is an observable of the space-time.

In several recent works \cite{Amorim1,Amorim4,Amorim5,Amorim2,EMCAbreuMJNeves2013}, a new version of NC
quantum mechanics (NCQM) were introduced, where not only
the coordinates ${\mathbf x}^\mu$ and their canonical momenta ${\mathbf p}_\mu$ are considered as operators in a Hilbert space ${\cal H}$, but also the objects of noncommutativity $\theta^{\mu\nu}$ and their canonical conjugate momenta $\pi_{\mu\nu}$.
All these operators belong to the same algebra and have the same hierarchical level, introducing  a minimal canonical
extension of DFR formalism. This enlargement of the usual
set of Hilbert space operators allows the theory to be invariant under the rotation group $SO(D)$, as showed
in detail in Ref. \cite{Amorim1,Amorim2}, when the treatment is a nonrelativistic one.  Rotation invariance in a nonrelativistic theory is fundamental if one intends to describe any physical system in a consistent way.  It was demonstrated in a precise way that in fact the DFR formalism has a momentum associated with $\theta^{\mu\nu}$.
In the present work we essentially consider the second quantization of the model discussed in Ref \cite{Amorim4}, showing that the extended Poincar\'e symmetry here is generated via generalized Heisenberg relations, giving the same algebra displayed in \cite{Amorim4,Amorim5}.

Although we have constructed a NC DFR Klein-Gordon equation \cite{aa} with a source term, an effective action using the Green functions was completely calculated in \cite{EMCAbreuMJNeves2012}. {The DFR NC model of scalar field with self-interaction $\phi^{4}$ was proposed in order to investigate the divergences at the one loop \cite{EMCAbreuMJNeves2014}.

The organization of the present paper follows that in section II we have described the DFR formalism.
In section III we have analyzed the charged Klein-Gordon and Dirac equations.  We have introduced new forms and constructions concerning the Gamma matrices in DFR formalism.  These new constructions complement those ones given in \cite{Amorim2}.  In section IV we have discussed of the Dirac action.   In section V we have dealt with the field equations and the invariants of the star-symmetry $U^{\star}(N)$ in DFR space.   To complete this analysis, in section 6, we have introduced the propagators of both gauge and fermionic fields.   the DFR diagrams for the propagators were  described in detail.
The conclusions and perspectives for future works, as always, are depicted in the final section.


\section{The DFR algebra in a nutshell}

In this section, we will review the main steps published in \cite{Amorim1,Amorim2,Amorim3,Amorim4,Amorim5}.
Namely, we will revisit the basics of the quantum field theory defined in the DFR space.
As we have said before, in DFR formalism the parameters $\theta^{\mu\nu}$ are promoted
to coordinate-operator in this space-time, which has $D=10$, it has six independents spatial coordinates,
which are, $\theta^{\mu\nu}=\left(\theta^{01},\theta^{02},\theta^{03},\theta^{12},\theta^{13},\theta^{23}\right)$.
Consequently, the coordinate $\theta^{\mu\nu}$ are promoted
to quantum observables $\hat{\theta}^{\mu\nu}$ in the commutation relation (\ref{xmuxnu}).
So we have the DFR algebra \cite{DFR1}
\begin{eqnarray}\label{algebraDFR}
\left[\hat{X}^{\mu},\hat{X}^{\nu}\right] = i \, \hat{\Theta}^{\mu\nu}
\hspace{0.2cm} , \hspace{0.2cm}
\left[\hat{x}^{\mu},\hat{\Theta}^{\nu\alpha}\right] = 0
\hspace{0.4cm} \mbox{and} \hspace{0.4cm}
\left[\hat{\Theta}^{\mu\nu},\hat{\Theta}^{\alpha\beta}\right] = 0 \; .
\end{eqnarray}
Moreover there exist the canonical conjugate momenta operator $\hat{K}^{\mu\nu}$ associated\footnote{The standard notation to represent the momentum is $\pi_{\mu\nu}$ but, for future ``to-avoid-confusion" convenience, we will use from now on, $\hat{K}^{\mu\nu}$.} with the operator $\hat{\Theta}^{\mu\nu}$, and they must satisfy the commutation relation
\begin{equation}\label{thetapicomm}
\left[\,\hat{\Theta}^{\mu\nu},\hat{K}^{\rho\sigma}\, \right] = i \, \uma^{\mu\nu\rho\sigma} \, \hat{\uma} \; ,
\end{equation}
where $\uma^{\mu\nu\rho\sigma}=\frac{1}{2}\left(\eta^{\mu\rho} \eta^{\nu\sigma} - \eta^{\mu\sigma} \eta^{\nu\rho} \right)$
is the anti-symmetrized identity matrix, and $\eta^{\mu\nu}=\mbox{diag}(1,-1,-1,-1)$ is the Minkowski metric.
In order to obtain consistency we can write that \cite{Amorim1}
\begin{eqnarray}\label{algebraDFRextended}
\left[\hat{X}^{\mu},\hat{P}^{\nu} \right] = i\,\eta^{\mu\nu} \, \hat{\uma}
\hspace{0.2cm} , \hspace{0.2cm}
\left[\hat{P}^{\mu},\hat{P}^{\nu} \right] = 0
\hspace{0.2cm} , \hspace{0.2cm}
\left[\hat{\Theta}^{\mu\nu},\hat{P}^{\rho}\right] = 0
\hspace{0.2cm} , \hspace{0.2cm}
\nonumber \\
\left[\hat{P}^{\mu},\hat{K}^{\nu\rho}\right] = 0
\hspace{0.2cm} , \hspace{0.2cm}
[\hat{X}^{\mu},\hat{K}^{\nu\rho}]=-{\frac i2}\,\uma^{\nu\rho\mu\sigma}\hat{P}_{\sigma} \; ,
\end{eqnarray}
and these relations complete the DFR extended algebra\footnote{Here we have adopted that $c=\hbar=\ell=1$, where the $\theta$-coordinate has area dimension.}.
It is possible to verify that (2.1)-(2.3) commutation relations listed above are indeed consistent with all possible
Jacobi identities and the CCR algebras \cite{EMCAbreuMJNeves2012}.

The uncertainty principle from (\ref{xmuxnu}) is modified by
\begin{eqnarray}\label{uncertainxmu}
\Delta \hat{X}^{\mu} \Delta \hat{X}^{\nu} \simeq \langle \hat{\Theta}^{\mu\nu}\rangle \; ,
\end{eqnarray}
where the expected value of the operator $\hat{\Theta}$ is related to the fluctuation position
of the particles, an it has dimension of length-squared.

The last commutation relation in Eq. (\ref{algebraDFRextended}) suggests that the shifted coordinate
operator \cite{Chaichian,Gamboa,Kokado,Kijanka,Calmet1,Calmet2}
\begin{equation}\label{X}
\hat{\xi}^{\mu} = \hat{X}^{\mu}\,+\,{\frac 12}\,\hat{\Theta}^{\mu\nu}\hat{P}_{\nu}\,\,,
\end{equation}
commutes with $\hat{K}^{\mu\nu}$.  The relation (\ref{X}) is also known in the algebraic literature as Bopp shift.
The commutation relation (\ref{algebraDFR}) also commutes with $\hat{\Theta}^{\mu\nu}$ and $\hat{\xi}^{\mu}$,
and satisfies a non trivial commutation relation with $\hat{P}^{\,\mu}$ dependent objects, which could be derived from
\begin{equation}\label{Xpcomm}
[\hat{\xi}^{\mu},\hat{P}^{\nu}]=i\,\eta^{\mu\nu} \, \hat{\uma}
\hspace{0.6cm} , \hspace{0.6cm}
[\hat{\xi}^{\mu},\hat{\xi}^{\nu}]=0\,\, ,
\end{equation}
and we can note that the property $\hat{P}_{\mu} \hat{\xi}^{\mu}=\hat{P}_{\mu} \hat{X}^{\mu}$ is easily verified.
Hence, we can see from these both equations that the shifted coordinated operator (\ref{X}) allows us to recover
the commutativity. The shifted coordinate operator $\hat{\xi}^{\mu}$ plays a fundamental role in NC
quantum mechanics defined in the $\left(x+\theta\right)$-space \cite{Amorim1}, since it is possible to form a basis with its eigenvalues.
So, differently from $\hat{X}^{\mu}$, we can say that $\hat{\xi}^{\mu}$ forms a basis in
Hilbert space. The framework showed above demonstrated that in NCQM, the physical coordinates
do not commute and the respective eigenvectors cannot be used to form a basis
in ${\cal H}={\cal H}_1 \oplus {\cal H}_2$ \cite{Amorim4}.  This can be accomplished
with the Bopp shift defined in (\ref{X}) with (\ref{Xpcomm}) as consequence.
%
%
%
%
%
%

%
The Lorentz generator group is
\begin{eqnarray}\label{Mmunu}
\hat{M}_{\mu\nu}=\,\hat{\xi}_{\mu}\hat{P}_{\nu}\,-\,\hat{\xi}_{\nu}\hat{P}_{\mu}
+\hat{\Theta}_{\nu\rho}\hat{K}^{\rho}_{\; \;\mu}-\hat{\Theta}_{\mu\rho}\hat{K}^{\rho}_{\; \;\nu} \; ,
\end{eqnarray}
and from (\ref{algebraDFRextended}) we can write the generators
for translations as $\hat{P}_{\mu} \rightarrow - i \partial_{\mu}\,\,$.
With these ingredients it is easy to construct the commutation relations
\begin{eqnarray}\label{algebraMP}
\left[ \hat{P}_\mu , \hat{P}_\nu \right] &=& 0
\hspace{0.2cm} , \nonumber \\
\left[ \hat{M}_{\mu\nu},\hat{P}_{\rho} \right] &=& \,i\,\big(\eta_{\mu\rho}\,\hat{P}_\nu
-\eta_{\nu\rho}\,\hat{P}_\mu\big) \; ,
\hspace{0.1cm} \nonumber \\
\left[\hat{M}_{\mu\nu} ,\hat{M}_{\rho\sigma} \right] &=& i\left(\eta_{\mu\sigma}\hat{M}_{\rho\nu}
-\eta_{\nu\sigma}\hat{M}_{\rho\mu}-\eta_{\mu\rho}\hat{M}_{\sigma\nu}+\eta_{\nu\rho}\hat{M}_{\sigma\mu}\right) \; ,
\end{eqnarray}
which closes the proper algebra.  We can say that $\hat{P}_\mu$ and $\hat{M}_{\mu\nu}$
are the DFR algebra generators.
An important point in DFR algebra issue is that the Weyl representation of NC operators obeying
the commutation relations keeps the usual form of the Moyal product. In this case, the Weyl map
is represented by
\begin{eqnarray}\label{mapweyl}
\hat{{\cal W}}(f)(\hat{X},\hat{\Theta})=\int\frac{d^{4}p}{(2\pi)^{4}}
\frac{d^{6}k}{(2\pi\lambda^{-2})^{6}} \;
\widetilde{f}(p,k) \; e^{ip_{\mu} \hat{X}^{\mu}+\frac{i}{2} \, k_{\mu\nu} \cdot \hat{\Theta}^{\mu\nu}} \; .
\end{eqnarray}
The Weyl symbol provides a map from the operator algebra to the functions algebra equipped with a star-product,
via the Weyl-Moyal correspondence
\begin{eqnarray}\label{WeylMoyal}
\hat{f}(\hat{X},\hat{\Theta}) \; \hat{g}(\hat{X},\hat{\Theta})
\hspace{0.3cm} \longleftrightarrow \hspace{0.3cm}
f(x,\theta) \star g(x,\theta) \; ,
\end{eqnarray}
where the star-product $\star$ is defined by
\begin{eqnarray}\label{ProductMoyal}
\left. f(x,\theta) \star g(x,\theta) =
e^{\frac{i}{2}\theta^{\mu\nu}\partial_{\mu}\partial^{\prime}_{\nu}}
f(x,\theta) \, \, g(x^{\prime},\theta) \right|_{x^{\prime}=x} \; ,
\end{eqnarray}
for any arbitrary functions $f$ and $g$ of the coordinates $(x^{\mu},\theta^{\mu\nu})$.
Namely, in both sides of Eq. (\ref{ProductMoyal}) we have that $f$ and $g$ are NC objects since they depend on $\theta^{\mu\nu}$.

The Weyl operator (\ref{mapweyl}) has the trace property considering a product of $n$ NC functions $(f_{1},...,f_{n})$
\begin{eqnarray}\label{traceWfs}
\mbox{Tr}\left[ \hat{{\cal W}}(f_{1}) ... \hat{{\cal W}}(f_{n}) \right]=
\int d^{4}x \; d^{6}\theta \; W(\theta) \; f_{1}(x,\theta) \star ... \star f_{n}(x,\theta) \; .
\end{eqnarray}

The function $W$ is a Lorentz invariant $\theta$-integration measure.
This weight function is introduced in the context of NC field theory to control divergences of the integration
in the $\theta$-space \cite{Carlson,Morita,Conroy2003,Saxell}. Theoretically speaking, it would permit us to work with series expansions
in $\theta$, {\it i.e.}, with truncated power series expansion of functions of $\theta$.
For any large $\theta^{\mu\nu}$ it falls to zero quickly so that all integrals are well defined,
in its definitions the normalization condition was assumed when integrated in the $\theta$-space.
The function $W$ should be an even function of $\theta$, that is, $W(-\theta)=W(\theta)$ which implies that
an integration in $\theta$-space be a Lorentz invariant. All the properties involving the
$W$-function can be seen in details in \cite{Carlson,Morita,Conroy2003,Saxell}.
However, we have to say that the role of the $W$-function in NC issues is not altogether clear among the NC community.
By the definition of the Moyal product (\ref{ProductMoyal}) it is
trivial to obtain the property
\begin{eqnarray}\label{Idmoyalproduct2}
\int d^{4}x\,d^{6}\theta \; W(\theta) \; f(x,\theta) \star g(x,\theta)=
\int d^{4}x\,d^{6}\theta \; W(\theta) \; f(x,\theta) \; g(x,\theta) \; .
\end{eqnarray}
%
%
%
%
The physical interpretation of the average of the components of $\theta^{\mu\nu}$, {\it i.e.}
$\langle \theta^{2} \rangle$, is the definition of the NC energy scale \cite{Carlson}
\begin{eqnarray}\label{LambdaNC}
\Lambda_{NC}=\left(\frac{12}{\langle \theta^{2} \rangle} \right)^{1/4}=\frac{1}{\lambda} \; ,
\end{eqnarray}
where $\lambda$ is the fundamental length scale
that appears in the Klein-Gordon (KG) equation (\ref{NCKG}) and in the dispersion relation
(\ref{RelDispDFR}) just below. This approach has the advantage of being unnecessary in order to
specify the form of the function $W$, at least for lowest-order processes.
The study of Lorentz-invariant NC QED,
as Bhabha scattering, dilepton and diphoton production
to LEP data led the authors of \cite{Conroy2003,Carone} to determine the bound
\begin{eqnarray}\label{boundLambda}
\Lambda_{NC} > 160 \; GeV \; \; 95 \% \; C.L. \; .
\end{eqnarray}

\section{Field theory in DFR space: Klein-Gordon and Dirac equations}

The first element of the algebra (\ref{algebraMP}) that commutes with all the others generators $\hat{P}^{\mu}$
and $\hat{M}^{\mu\nu}$ is given by $\hat{C}_{1}=\hat{P}_{\mu}\hat{P}^{\mu}+\frac{\lambda^{2}}{2}\hat{K}_{\mu\nu}\hat{K}^{\mu\nu}$.
This is the first Casimir operator of the algebra (\ref{algebraMP}). Using the coordinate representation, the operators
$\hat{P}^{\mu}$ and $\hat{K}^{\mu\nu}$ can be written in terms of the derivatives
\begin{eqnarray}\label{repcoordinateppi}
\hat{P}_{\mu} \longmapsto -i\p_{\mu}
\hspace{0.6cm} \mbox{and} \hspace{0.6cm}
\hat{K}_{\mu\nu} \longmapsto -i\frac{\p}{\p \theta^{\mu\nu}} \; ,
\end{eqnarray}
and consequently, the first Casimir operator in the on-shell condition leads
us to the KG equation in DFR space concerning the scalar field $\phi$
\begin{eqnarray}\label{NCKG}
\left(\Box +\lambda^2\Box_\theta+m^2\right)\phi(x,\theta)=0 \; ,
\end{eqnarray}
where we have defined $\Box_{\theta}=\frac{1}{2}\,\p^{\mu\nu}\,\p_{\mu\nu}$
and $\p_{\mu\nu}:=\frac{\p}{\p \theta^{\mu\nu}}$. The plane wave general solution
for the DFR KG equation is the Fourier integral
\begin{eqnarray}\label{phiXtheta}
\phi(x,\theta)=\int \frac{d^{4}p}{(2\pi)^{4}} \frac{d^{6}k}{(2\pi\lambda^{-2})^{6}}
\, \widetilde{\phi}(p,k^{\mu\nu}) \, \exp \left( ip_{\mu}x^{\mu}+\frac{i}{2} k_{\mu\nu} \theta^{\mu\nu} \right) \; .
\end{eqnarray}
%
The length $\lambda^{-2}$ is introduced
conveniently in the $k$-integration to maintain the field dimension
as being length inverse. Consequently, the $k$-integration keeps dimensionless.
Substituting the wave plane solution (\ref{phiXtheta}), we obtain the invariant mass
\begin{eqnarray}\label{MassInv}
p^{2}+\frac{\lambda^{2}}{2}k_{\mu\nu}k^{\mu\nu}=m^2 \; ,
\end{eqnarray}
where $\lambda$ is the parameter with length dimension defined before, it
is a Planck-type length. We can define the components of the $k$-momentum
$k^{\mu\nu}=(-{\bf k},-\widetilde{{\bf k}})$ and $k_{\mu\nu}=({\bf k},\widetilde{{\bf k}})$,
to obtain the DFR dispersion relation
\begin{eqnarray}\label{RelDispDFR}
\omega({\bf p},{\bf k},\widetilde{{\bf k}})=\sqrt{{\bf p}^{2}
+\lambda^{2}\left({\bf k}^{2}+\widetilde{{\bf k}}^{2}\right)+m^2} \; ,
\end{eqnarray}
where $\widetilde{k}_{i}$ is the vector connected to the components $k_{ij}$,
that is, $k_{ij}=\epsilon_{ijk}\widetilde{k}_{k}$ $(i,j,k=1,2,3)$.
It is easy to see that, using the limit $\lambda \rightarrow 0$ in Eqs.
(\ref{NCKG})-(\ref{RelDispDFR}) we can recover the commutative expression
\cite{EMCAbreuMJNeves2012}.
Since we have constructed the NC KG equation, we will now show its relative action.
We will use the definition of the Moyal-product (\ref{ProductMoyal}) to write the action
for a free complex scalar field in DFR scenario as being
\begin{equation}\label{actionscalarstar}
S_{KG}(\phi^{\ast},\phi)=\int d^{4}x \,d^{6}\theta \, W(\theta) \left( \partial_{\mu}\phi^{\ast} \star \partial^{\mu}\phi
+ {{\lambda^2}\over 2} \, \partial_{\mu\nu}\phi^{\ast} \star \partial^{\mu\nu}\phi
-m^2\phi^{\ast} \star \phi \right) \; ,
\end{equation}
and using the identity (\ref{Idmoyalproduct2}), this free action
can be reduced to the usual one
\begin{eqnarray}\label{actionscalar}
S_{KG}(\phi^{\ast},\phi)=\int d^{4}x \,d^{6}\theta \, W(\theta) \left( \left|\partial_{\mu}\phi\right|^{2} +
{{\lambda^2}\over 2} \left|\partial_{\mu\nu}\phi\right|^{2}
-m^2\left|\phi\right|^{2} \right) \; .
\end{eqnarray}
\ni where all the products are the usual ones.

It is easy to show that the DFR Dirac equation can be deduced from the square root of the DFR KG equation,
so we can write the field equation \cite{Amorim5}
\begin{eqnarray}\label{EqDiracDFR}
\Big(\,i\gamma^{\mu}\partial_{\mu}
+\,{{i\lambda}\over2}\,\Gamma^{\mu\nu}\partial_{\mu\nu}
-m \Big) \psi(x,\theta)=0 \; ,
\end{eqnarray}
where $\gamma^{\mu}$'s are the ordinary Dirac matrices, and they satisfy the usual relations
\begin{eqnarray}\label{gammamu}
\left\{ \gamma^{\mu},\gamma^{\nu} \right\}=2\,\eta^{\mu\nu} \; .
\end{eqnarray}
The matrices $\Gamma^{\mu\nu}$ are six matrices $4\times4$ which, by construction, they must be anti-symmetric, i.e., $\Gamma^{\mu\nu}=-\Gamma^{\nu\mu}$.
We can write the matrices $\Gamma^{\mu\nu}$ in terms of Dirac matrices-$\gamma$ by means of the commutation relation
\begin{eqnarray}
\Gamma^{\mu\nu}:=\frac{i}{2}\left[\,\gamma^{\mu},\gamma^{\nu}\,\right] \; ,
\end{eqnarray}
where we can show that the anti-commutation relations are given by
\begin{eqnarray}\label{gammamunu}
\left\{\gamma^{\mu},\Gamma^{\nu\rho} \right\}=i\eta^{\mu\rho}\gamma^{\nu}-i\eta^{\mu\nu}\gamma^{\rho}
+\Gamma^{\mu\nu}\gamma^{\rho}-\Gamma^{\mu\rho}\gamma^{\nu}
\hspace{0.4cm} \mbox{and}
\nonumber \\
\left \{ \Gamma^{\mu\nu},\Gamma^{\rho\lambda} \right\}= \gamma^{\rho}\gamma^{\mu}\eta^{\nu\lambda}+\gamma^{\lambda}\gamma^{\nu}\eta^{\mu\rho}
-\gamma^{\lambda}\gamma^{\mu}\eta^{\nu\rho}
-\gamma^{\rho}\gamma^{\nu}\eta^{\mu\lambda} \; ,
\hspace{1.5cm}
\end{eqnarray}
and the hermiticity property of $\Gamma^{\mu\nu}$ is the same as $\gamma^{\mu}$, i.e., $\left(\Gamma^{\mu\nu}\right)^{\dagger}=\gamma^{0}\Gamma^{\mu\nu}\gamma^{0}$.
Using these relations, the Dirac equation (\ref{EqDiracDFR}) leads us to DFR Klein-Gordon equation.
The components of $\Gamma^{\mu\nu}=\left(\Gamma^{0i},\Gamma^{ij}\right)$ can be written in terms of the Pauli matrices as
\begin{eqnarray}
\Gamma^{0i}=i\left(
  \begin{array}{cc}
    0 & \sigma^{i} \\
    \sigma^{i} & 0 \\
  \end{array}
\right)
\hspace{0.3cm} , \hspace{0.3cm}
\Gamma^{ij}=\left(
  \begin{array}{cc}
    \epsilon^{ijk}\sigma^{k} & 0 \\
    0 & \epsilon^{ijk}\sigma^{k} \\
  \end{array}
\right) \; \; , \; \; i,j=1,2,3 \; .
\end{eqnarray}
which complements the results obtained in \cite{Amorim5}.

It can be shown that the connection between the Dirac equation and its adjoint equation can lead us to a conservation law
\begin{equation}\label{conservlei}
\partial_{\mu}J^{\mu}+\frac{\lambda}{2} \, \partial_{\mu\nu}{\cal J}^{\mu\nu}=0 \; ,
\end{equation}
where $J^{\mu}:=\bar{\psi}\, \gamma^{\mu} \, \star \, \psi$, and ${\cal J}^{\mu\nu}:=\bar{\psi}\, \Gamma^{\mu\nu} \, \star \psi$
are the currents that emerge from the DFR Dirac equation.
By integrating the expression (\ref{conservlei}) considering the whole space ($x+\theta$), the Dirac field charge $\psi^{\dagger} \psi$
is conserved, as the commutative usual case.  Notice that the new current term $\bar{\psi}\, \Gamma^{\mu\nu} \, \star \psi$ has the generator of the
rotational group attached to it. In the next section we will investigate the coupling of this current with the gauge fields,
which can be an interesting study of the Yang-Mills model in DFR phase-space.
The DFR action for the Dirac field is
\begin{equation}\label{actionDiracDFRstar}
S_{Dirac}(\bar{\psi},\psi)=\int d^{4}x\,d^{6}\theta\, W(\theta) \, \bar{\psi}(x,\theta) \star \Big(\,i\gamma^{\mu}\partial_{\mu}
+\,{{i\lambda}\over2}\,\Gamma^{\mu\nu}\partial_{\mu\nu}
-m \Big) \psi(x,\theta) \, \,,
\end{equation}
which, using the identity (\ref{ProductMoyal}) can be reduced to
\begin{equation}\label{actionDiracDFRstar2}
S_{Dirac}(\bar{\psi},\psi)=\int d^{4}x\,d^{6}\theta\, W(\theta) \, \bar{\psi}(x,\theta)
\Big(\,i\gamma^{\mu}\partial_{\mu}
+\,{{i\lambda}\over2}\,\Gamma^{\mu\nu}\partial_{\mu\nu}
-m \Big) \psi(x,\theta) \, \,,
\end{equation}
which is invariant by symmetry transformations of the Poincar\'e DFR algebra \cite{Amorim5}.
In the next section we will discuss the gauge symmetries of the DFR Dirac Lagrangian.

%

\section{Gauge symmetry and the $U^{\star}(N)$ action}

The actions of the complex scalar and Dirac field are invariant under
global transformations of the fields. The invariance of the action under
this global symmetry gives rise to conserved charges, indicated by the conservation law
(\ref{conservlei}). We will now discuss in the invariance of the Dirac action under
local transformations. Let us consider the local gauge transformations for the spinors fields
\begin{eqnarray}\label{Localtransfpsi}
\Psi(x,\theta) \hspace{0.2cm} \longmapsto \hspace{0.2cm}
\Psi^{\prime}(x,\theta)=U(x,\theta)\star\Psi(x,\theta) \; ,
\end{eqnarray}
where $U(x,\theta)$ is an arbitrary matrix $N \times N$ of the coordinates $(x,\theta)$.
It must satisfy the unitarity property
\begin{eqnarray}\label{unitaryU}
U(x,\theta) \star U^{\dagger}(x,\theta)=U^{\dagger}(x,\theta) \star U(x,\theta)=\uma_{N} \; ,
\end{eqnarray}
and we can say that $U$ is {\it star-unitary}, and $\uma_{N}$ is the identity matrix $N \times N$.
%
%
The DFR Dirac Lagrangian is not invariant under the local transformation (\ref{Localtransfpsi}).
To obtain such invariance, we must replace both derivatives $\partial_{\mu}$ and $\lambda\partial_{\mu\nu}$
by the following covariant derivatives
\begin{eqnarray}\label{DmuDmunu}
\partial_{\mu} \longmapsto D_{\mu} \star= \partial_{\mu} + i g A_{\mu} \star
\nonumber \\
\lambda \, \partial_{\mu\nu} \longmapsto D_{\mu\nu} \star= \lambda \, \partial_{\mu\nu} + i g^{\prime} B_{\mu\nu} \star \; ,
\end{eqnarray}
and the NC Dirac Lagrangian is
%
%
%
%
\begin{eqnarray}\label{LDiracDcov}
{\cal L}_{Dirac-gauge}=\bar{\Psi}(x,\theta) \star \Big(\,i\gamma^{\mu}D_{\mu}\star
+\,{{i}\over2}\,\Gamma^{\mu\nu}D_{\mu\nu}\star
-m \Big) \Psi(x,\theta)  \; .
\end{eqnarray}
The first one is the usual covariant derivative with a star-product, while $D_{\mu\nu}\star$ is a
new anti-symmetric star-covariant derivative associated with the $\theta$-space
%
%
Consequently, the new field $B_{\mu\nu}$ is an anti-symmetric tensor field $(B_{\mu\nu}=-B_{\nu\mu})$.
It has six independent components,
{\it i.e.} $B^{\mu\nu}=\left(B^{0i},B^{ij}\right)$, where $i,j=\{1,2,3\}$,
which defines the tensor field in the $\left(x+\theta\right)$-space.
The notation $D_{\mu}\star$ indicates a star-product between $A_{\mu}$
and the Dirac spinor, which does  also occur within the covariant derivative $D_{\mu\nu}\star$ case.
For convenience we have introduced the coupling constants $g$, and $g^{\prime}$
associated with the $\theta$-space. To complete such invariance, we must impose the DFR {\it star-gauge transformations}
\begin{eqnarray}\label{ABgaugetransf}
A_{\mu}(x,\theta) \,\, \longmapsto \,\, A_{\mu}^{\prime}(x,\theta)=U(x,\theta)\star A_{\mu}(x,\theta)\star U^{\dagger}(x,\theta)-\frac{i}{g}\left(\partial_{\mu}U\right) \star U^{\dagger}(x,\theta) \; ,
\nonumber \\
B_{\mu\nu}(x,\theta) \,\, \longmapsto \,\, B_{\mu\nu}^{\prime}(x,\theta)=U(x,\theta) \star B_{\mu\nu}(x,\theta) \star U^{\dagger}(x,\theta)-\frac{i}{g^{\prime}}\left(\lambda\partial_{\mu\nu}U\right) \star U^{\dagger}(x,\theta) \; .
\hspace{0.6cm}
\end{eqnarray}

Note that (\ref{unitaryU}) implies that $U^{\dagger}$ is equal to $U^{-1}$ with respect to the star-product
upon the deformed algebra of functions on space-time. In general, it is not true for $\theta \neq 0$,
in which $U^{\dagger} \neq U^{-1}$. An explicit relation between both $U^{\dagger}$ and $U^{-1}$ can be obtained by the series
of the star product, that can be written as
\begin{eqnarray}
U^{\dagger}=U^{-1}+\frac{i}{2} \, \theta^{\mu\nu} \, U^{-1} \, \partial_{\mu}U \, U^{-1} \, \partial_{\nu}U \, U^{-1}
+{\cal O}(\theta^{2}) \; .
\end{eqnarray}
Due to the property $\left(f \star g\right)^{\dagger}=g^{\dagger} \star f^{\dagger}$, the Moyal product $f \star g$
of two unitary matrix fields is always unitary and the group $U^{\star}(N)$ is closed under the star-product.
The special unitary group $SU(N)$ does not give rise to any gauge group in the NC space-time,
because in general $\mbox{det}(f\star g) \neq \mbox{det}(f) \star \mbox{det}(g)$, and consequently,
$\mbox{det}(U \, \star \, U^{\dagger}) \neq \mbox{det}(U) \, \star \, \mbox{det}(U^{\dagger})$, that is, $\det U \neq 1$.

In the opposite case, relative to the
commutative case, $U^{\star}_{N}(1)$ and $SU^{\star}(N)$ are sectors of the decomposition $U^{\star}(N)=U_{N}^{\star}(1) \times SU^{\star}(N)$
do not decouple because the gauge fields of $U_{N}^{\star}(1)$ interacts with the gauge fields of $SU^{\star}(N)$. We represent the $U$-function of $U^{\star}(N)$ as the $\star$-product
\begin{equation}\label{Uexp}
U(x,\theta)=e^{i\alpha(x,\theta)\,\uma_{N}} \star e^{it^{a}\omega^{a}(x,\theta)} \; ,
\end{equation}
where $\alpha$, $\omega^{a}$ are arbitrary real functions of $(x,\theta)$ associated with the NC Abelian subgroup
$U_{N}^{\star}(1)$ and with the NC non-Abelian subgroup $SU^{\star}(N)$, respectively, and $t^{a} \, (a=1,2,\cdots, N^{2}-1)$
are the traceless generators
of the Lie algebra of the subgroup $SU^{\star}(N)$. The fields $\left(A_{\mu},B_{\mu\nu}\right)$ are hermitian gauge fields
of the star unitary group $U_{\star}(N)$ defined in the DFR NC space-time scenario. They can be expanded in terms of the Lie algebra
generators of $U_{\star}(N)$ as $A_{\mu}=A_{\mu}^{0} \, {\uma_{N}} +A_{\mu}^{a}t^{a}$ and
$B_{\mu\nu}=B_{\mu\nu}^{0} \, {\uma_{N}} +B_{\mu\nu}^{a}t^{a}$,
with $\mbox{tr}_{N}(t^{a}t^{b})=\delta^{ab}$, $a,b=1,...,N^{2}-1$, by obeying the Lie algebra commutation
relation $\left[t^{a},t^{b}\right]=if^{abc}t^{c}$, and the anti-commutation algebra $\left\{t^{a},t^{b}\right\}=d^{abc}t^{c}$.
The constants $f^{abc}$ and $d^{abc}$ are the structure constants of the Lie algebra. The fields $A_{\mu}^{0}$ and $B_{\mu\nu}^{0}$
come from the Abelian part of the group $U^{\star}(N)$, while the components $A_{\mu}^{a}$ and $B_{\mu\nu}^{a}$ are attached
to the non-Abelian part of $U^{\star}(N)$.
Here, the generators $t^{a}$ live in the adjoint representation of the $U^{\star}(N)$ gauge group, and $\mbox{tr}_{N}$ denotes the matrix
trace.  Notice that in the fermionic sector, the spinor field is the column matrix  whose components are
$\Psi=\left(\psi_{1},\psi_{2}, \cdot \cdot \cdot, \psi_{N}\right)$ that live in the basic
representation of the Lie algebra. An important issue is that expressions in NC
gauge theory involve the enveloping algebra of the underlying Lie Group. The components $A_{\mu}^{\,0}$ and $A_{\mu}^{\,a}$
have the infinitesimal gauge transformation from (\ref{ABgaugetransf})
\begin{eqnarray}
A_{\mu}^{\prime \,\, 0}&=&A_{\mu}^{0}+i\left[\alpha(x,\theta),A_{\mu}^{0}\right]_{\star}+g^{-1}\partial_{\mu}\alpha(x,\theta) \; ,
\nonumber \\
\nonumber \\
A_{\mu}^{\prime\,\,a}&=&A_{\mu}^{a}-\left[\,\omega(x,\theta),A_{\mu}\right]_{\star}^{a}
+i\left[\alpha(x,\theta),A_{\mu}^{a} \right]_{\star}+i\left[\omega^{a}(x,\theta),A_{\mu}^{0} \right]_{\star}
+g^{-1}\partial_{\mu}\omega^{a}(x,\theta) \; ,
\end{eqnarray}
and for $B_{\mu\nu}$, we can have that
\begin{eqnarray}
B_{\mu\nu}^{\prime \, \,0} &=& B_{\mu\nu}^{0}+i\left[\alpha(x,\theta),B_{\mu\nu}^{0}\right]_{\star}+g^{\prime -1}\lambda\partial_{\mu\nu}\alpha(x,\theta) \; ,
\nonumber \\
\nonumber \\
B_{\mu\nu}^{\prime \, \, a}&=&B_{\mu\nu}^{a}-\left[\,\omega(x,\theta),B_{\mu\nu}\right]_{\star}^{a}
+i\left[\alpha(x,\theta),B_{\mu\nu}^{a} \right]_{\star}+i\left[\omega^{a}(x,\theta),B_{\mu\nu}^{0} \right]_{\star}
+g^{\prime \, -1}\lambda\partial_{\mu\nu}\omega^{a}(x,\theta) \; .
\nonumber \\
\end{eqnarray}
Using the Moyal product properties, the commutator $\left[\,\omega,A_{\mu}\right]_{\star}^{a}$ is given by the combination
of the cosine and sine series
\begin{eqnarray}
\label{commutatoromegaA}
\left. \left[\,\omega(x,\theta),A_{\mu}(x,\theta)\right]_{\star}^{a}= f^{abc} \, \cos\left(\frac{\theta^{\alpha\beta}}{2}\,\partial_{\alpha}\partial_{\beta}^{\prime} \right)
\omega^{b}(x,\theta) A_{\mu}^{c}(x^{\prime},\theta) \right|_{x^{\prime}=x}
\nonumber \\
\left. + d^{abc} \, \sin\left(\frac{\theta^{\alpha\beta}}{2}\,\partial_{\alpha}\partial_{\beta}^{\prime} \right)
\omega^{b}(x,\theta) A_{\mu}^{c}(x^{\prime},\theta) \right|_{x^{\prime}=x} \; ,
\end{eqnarray}
and the analogous to the case of $\left[\,\omega , B_{\mu\nu}\right]_{\star}^{a}$.  The simplest commutator
$\left[\alpha , A_{\mu}^{\; 0} \right]_{\star}$ is just the trigonometric sine part of (\ref{commutatoromegaA}),
with $f^{abc}=0$ and $d^{abc}=1$, that goes to zero in the commutative limit.

The $F_{\mu\nu}$ tensor associated with the $A_{\mu}$ field is defined as the {\it star commutation relation}
\begin{eqnarray}\label{DmuDnuFmunu}
\left[D_{\mu},D_{\nu}\right]_{\star}=ig \, F_{\mu\nu} \; ,
\end{eqnarray}
where
\begin{eqnarray}\label{Fmunu}
F_{\mu\nu}=\partial_{\mu}A_{\nu}-\partial_{\nu}A_{\mu}+i g \, \left[A_{\mu},A_{\nu}\right]_{\star} \; ,
\end{eqnarray}
and it has the gauge transformation
\begin{eqnarray}\label{transfgaugeFmunu}
F_{\mu\nu} \; \longmapsto \; F_{\mu\nu}^{\;\prime} = U(x,\theta) \star F_{\mu\nu} \star U^{\dagger}(x,\theta) \; .
\end{eqnarray}
It is easy to verify that the $F_{\mu\nu}$ tensor is Lie algebra valued of $U^{\star}(N)$ as $F_{\mu\nu}=F_{\mu\nu}^{\,0}{\uma_{N}}+F_{\mu\nu}^{a}\,t^{a}$, where the components are given by
%
\begin{eqnarray}\label{ExpF}
F_{\mu\nu}^{0}&=&\partial_{\mu}A_{\nu}^{0}-\partial_{\nu}A_{\mu}^{0}+ig\left[A_{\mu}^{0},A_{\nu}^{0} \right]_{\star}
\; ,
\nonumber \\
F_{\mu\nu}^{a}&=&\partial_{\mu}A_{\nu}^{a}-\partial_{\nu}A_{\mu}^{a}
-\frac{1}{2} \, g \, f^{abc} \,
\left\{A_{\mu}^{b}, A_{\nu}^{c}\right\}_{\star} \!
\nonumber \\
&&+\,ig\left[A_{\mu}^{0},A_{\nu}^{a} \right]_{\star}
\!\!+ig\left[A_{\mu}^{a},A_{\nu}^{0} \right]_{\star}
\!\!+\frac{i}{2} \, g \, d^{abc} \,
\left[A_{\mu}^{b}, A_{\nu}^{c}\right]_{\star} \; .
\end{eqnarray}
These components can be understood as the NC electromagnetic field tensor $F_{\mu\nu}^{0}$, and the Yang-Mills tensor
$F_{\mu\nu}^{a}$ defined in the NC space-time DFR.
%
%
%
Analogously, we obtain the field strength tensor of $B_{\mu\nu}$
by calculating the star commutation relation
\begin{eqnarray}\label{RelCommutDmunuG}
\left[D_{\mu\nu} , D_{\sigma\rho} \right]_{\star}=ig^{\prime} \, G_{\mu\nu\rho\sigma} \; ,
\end{eqnarray}
where the components of $G$ are
\begin{eqnarray}\label{Gmunu}
G_{\mu\nu\rho\sigma}^{0}&=&\lambda\partial_{\mu\nu}B_{\rho\sigma}^{0}
-\lambda\partial_{\rho\sigma}B_{\mu\nu}^{0}+ig^{\prime}\left[B_{\mu\nu}^{0} , B_{\rho\sigma}^{0}\right]_{\star}
\nonumber \\
G_{\mu\nu\rho\sigma}^{a}&=&\lambda\partial_{\mu\nu}B_{\rho\sigma}^{a}
-\lambda\partial_{\rho\sigma}B_{\mu\nu}^{a}
- \frac{1}{2} \, g^{\prime} \, f^{abc} \,
\left\{B_{\mu\nu}^{b}, B_{\rho\sigma}^{c}\right\}_{\star}
\nonumber \\
&&+ig^{\prime}\left[B_{\mu\nu}^{0},B_{\rho\sigma}^{a} \right]_{\star}
\!\!+ig^{\prime}\left[B_{\mu\nu}^{a},B_{\rho\sigma}^{0} \right]_{\star}
+\frac{i}{2} \, g^{\prime} \, d^{abc} \,
\left[B_{\mu\nu}^{b} , B_{\rho\sigma}^{c} \right]_{\star}  .
\end{eqnarray}
%
%
It has the anti-symmetric properties
\begin{eqnarray}\label{propertiesG}
G_{\mu\nu\rho\sigma}=-G_{\nu\mu\rho\sigma}
=-G_{\mu\nu\sigma\rho}=G_{\nu\mu\sigma\rho}
=-G_{\rho\sigma\mu\nu} \; ,
\end{eqnarray}
and the gauge transformation
\begin{eqnarray}\label{transfgaugeG}
G_{\mu\nu\rho\sigma} \; \longmapsto \;
G^{\;\prime}_{\mu\nu\rho\sigma}=U(x,\theta) \star G_{\mu\nu\rho\sigma} \star U^{\dagger}(x,\theta) \; .
\end{eqnarray}
From (\ref{transfgaugeFmunu}) and (\ref{transfgaugeG}), the infinitesimal transformations of the components of $F_{\mu\nu}$ and $G_{\mu\nu\rho\sigma}$ are given by
\begin{eqnarray}\label{TransfinfFG}
F_{\mu\nu}^{0} \; \longmapsto \; F_{\mu\nu}^{0 \, \prime} &=& F_{\mu\nu}^{0}+i \, [\alpha,F_{\mu\nu}^{0}]_{\star} \; ,
\nonumber \\
F_{\mu\nu}^{a} \; \longmapsto \; F_{\mu\nu}^{a \, \prime} &=& F_{\mu\nu}^{a}-\frac{1}{2}\, f^{abc} \left\{\omega^{b},F_{\mu\nu}^{c} \right\}_{\star}
+i \, \left[\alpha,F_{\mu\nu}^{a}\right]_{\star}+
\nonumber \\
&&+i \, \left[\omega^{a},F_{\mu\nu}^{0}\right]_{\star}
+i \, d^{abc}\left[\omega^{b},F_{\mu\nu}^{c} \right]_{\star} \; ,
\nonumber \\
G_{\mu\nu\rho\sigma}^{0} \; \longmapsto \; G_{\mu\nu\rho\sigma}^{0 \, \prime} &=& G_{\mu\nu\rho\sigma}^{0}+i \, [\alpha,G_{\mu\nu\rho\sigma}^{0}]_{\star} \; ,
\nonumber \\
G_{\mu\nu\rho\sigma}^{a} \; \longmapsto \; G_{\mu\nu\rho\sigma}^{a \, \prime} &=& G_{\mu\nu\rho\sigma}^{a}-\frac{1}{2}\, f^{abc} \left\{\omega^{b},G_{\mu\nu\rho\sigma}^{c} \right\}_{\star}+i \, \left[\alpha,G_{\mu\nu\rho\sigma}^{a}\right]_{\star}+
\nonumber \\
&&+i \, \left[\omega^{a},G_{\mu\nu\rho\sigma}^{0}\right]_{\star}
+i \, d^{abc}\left[\omega^{b},G_{\mu\nu\rho\sigma}^{c} \right]_{\star} \; .
\end{eqnarray}
%
%
%
Therefore we have a Lagrangian for the gauge fields which is invariant under the transformations
(\ref{transfgaugeFmunu}) and (\ref{transfgaugeG})
\begin{equation}\label{LAB}
{\cal L}_{gauge}=
-\frac{1}{4}\,\mbox{tr}_{N}\left(F_{\mu\nu}\star F^{\mu\nu}\right)
-\frac{1}{4}\,\mbox{tr}_{N}\left(G_{\mu\nu\rho\sigma}\star
G^{\mu\nu\rho\sigma}\right)
-\frac{1}{2}\,\mbox{tr}_{N}\left(F_{\mu\nu} \star G^{\mu\rho\nu}_{\hspace{0.5cm}\rho}
\right)
\; .
\end{equation}
Analogously, this invariance can be applied to the KG Lagrangian of (\ref{actionscalarstar})
by substituting the ordinary derivatives $\partial_{\mu}$ and $\partial_{\mu\nu}$,
by the covariant derivatives $D_{\mu}$ and $D_{\mu\nu}$, respectively, we can write
\begin{eqnarray}\label{LescalarDDmunu}
{\cal L}_{KG-gauge}=\mbox{tr}_{N}\left(D_\mu\Phi^{\dagger} \star D^\mu\Phi\right) +
\frac{1}{2} \,\mbox{tr}_{N}\left(D_{\mu\nu}\Phi^{\dagger} \star D^{\mu\nu}\Phi\right)
-m^2\,\mbox{tr}_{N}\left(\Phi^{\dagger} \star \Phi\right) \; ,
\end{eqnarray}
where the scalar field $\Phi$ represents a multiplet of $N$ complex scalar fields, namely,
$\Phi=(\phi_{1},\phi_{2},\cdots,\phi_{N})$.

The DFR version of a quantum electrodynamics (QED) is represented by the group $U^{\star}(1)$,
with $N=1$, where we have just one field $A^{\mu}$ and the anti-symmetric $B^{\mu\nu}$. The gauge
transformations are analogous to (\ref{ABgaugetransf}) and (\ref{transfgaugeFmunu}), with
$U(x,\theta)=e^{i \, {\uma} \, \alpha(x,\theta)}$, which obeys the Moyal product series
\begin{equation}\label{Eseries}
U(x,\theta)=1+i\alpha(x,\theta)+\frac{i^{2}}{2!} \, \, \alpha(x,\theta) \star \alpha(x,\theta) + {\cal O}(\alpha^{3}) \; .
\end{equation}
The expressions of the DFR electromagnetic tensors can be obtained by making both $f^{abc}=0$ and $d^{abc}=1$ in (\ref{ExpF}) and (\ref{Gmunu}).  By
redefining $A_{\mu}^{0}=A_{\mu}$, $B_{\mu\nu}^{0}=B_{\mu\nu}$ and the coupling constants $g=e$,
$g^{\prime}=e^{\prime}$, we obtain
\begin{eqnarray}\label{ExpFem}
F_{\mu\nu}&=&\partial_{\mu}A_{\nu}-\partial_{\nu}A_{\mu}
+ ie \, \left[A_{\mu} , A_{\nu}\right]_{\star} \; \; \; ,
\nonumber \\
\nonumber \\
G_{\mu\nu\rho\sigma}&=&\lambda\partial_{\mu\nu}B_{\rho\sigma}
-\lambda\partial_{\rho\sigma}B_{\mu\nu}
+ ie^{\prime} \, \left[B_{\mu\nu} , B_{\rho\sigma}\right]_{\star} \;  ,
\end{eqnarray}
%
%
%
%
where we have achieved an invariance property for the DFR Dirac action under the local transformations
(\ref{Localtransfpsi}). To guarantee this invariance we must introduce an anti-symmetric
field $B_{\mu\nu}$ (\ref{DmuDmunu}), beyond the vector field $A_{\mu}$,
since these fields have the gauge transformations (\ref{ABgaugetransf}).
This new gauge anti-symmetric field is associated with the $\theta$-space
and it must to be attached to those extra-dimensions.
The NC gauge theory obtained here is reduced to the standard case of $SU(N)$ Yang-Mills,
and the usual $U(1)$ QED, in the commutative limit $\theta=0$, and taking $\lambda=0$ in the Lagrangian (\ref{LDiracDcov}),
(\ref{LAB}) and (\ref{LescalarDDmunu}).

In the next section we will discuss the field equations and the currents of the star-symmetry $U^{\star}(N)$ in DFR space.
%
%


%
\section{Field equations and the DFR Electromagnetism}

The electromagnetic and Yang-Mills field equations in DFR space-time will
be computed in this section. To accomplish the task, we have to find both the Dirac and gauge Lagrangian, Eq. (\ref{LDiracDcov}) and (\ref{LAB}) respectively, which obey the star gauge symmetry $U^{\star}(N)$ discussed in the last section
\begin{eqnarray}\label{LSpinor+gauge}
{\cal L}_{U^{\star}(N)}&=&\bar{\psi}_{i} \, \star \, \Big(\,i\gamma^{\mu}D_{\mu}\star
+\,\frac{i}{2}\,\Gamma^{\mu\nu}D_{\mu\nu}\star-m \Big)_{ij} \psi_{j}
-\frac{1}{4}\,F_{0\mu\nu} \, \star \, F_{0}^{ \; \mu\nu}
-\frac{1}{4}\, G_{0\mu\nu\rho\sigma}
\, \star \, G_{0}^{\mu\nu\rho\sigma}
\nonumber \\
&&-\frac{1}{2}\,F_{0\mu\nu} \, \star \, G^{ \hspace{0.15cm} \mu\rho\nu}_{0\hspace{0.44cm}\rho}
\,-\frac{1}{4}\,F_{\mu\nu}^{\hspace{0.3cm} a} \, \star \, F^{\mu\nu a}
-\frac{1}{4}\, G_{\mu\nu\rho\sigma}^{\hspace{0.68cm} a}
\, \star \, G^{\mu\nu\rho\sigma a}
-\frac{1}{2}\,F_{\mu\nu}^{\hspace{0.3cm} a} \, \star \, G^{\mu\rho\nu \hspace{0.2cm} a}_{\hspace{0.45cm}\rho} \; ,
\nonumber \\
\end{eqnarray}
where we have calculated the traces present in Eq. (\ref{LAB}). In the sector of the gauge fields we have naturally the
NC Maxwell Lagrangian and the Lagrangian of the NC Yang-Mills field. We treat
$\left(A_{0}^{\,\mu},A^{\mu a}\right)$ and $\left(B_{0}^{\mu\nu},B^{\mu\nu a}\right)$ as independent fields
to obtain the NC field equations. Using the principle of the minimal action associated with the Lagrangian
(\ref{LSpinor+gauge}), with respect to $A_{0}^{\mu}$, we can obtain the NC Maxwell's equations
in the presence of a source
\begin{eqnarray}\label{EqfieldA0}
\nabla_{0\mu} \, \star \, F_{0}^{\,\mu\nu}
+\nabla_{0\mu} \, \star \, G^{\hspace{0.16cm} \mu\rho\nu}_{0\hspace{0.45cm}\rho} =g \, J_{0}^{\; \, \nu} \; ,
\end{eqnarray}
where the covariant derivative $\nabla_{0\mu}$ acting on strength field tensors is defined by
\begin{eqnarray}\label{nabla0}
\nabla_{0\mu} \star F_{0}^{\; \mu\nu}:=\partial_{\mu}F_{0}^{\; \mu\nu}-g\left[A_{0 \, \mu},F_{0}^{ \; \mu\nu}\right]_{\star} \; .
\end{eqnarray}
The tensor $F_{0\mu\nu}$ must obey the Bianchi identity
\begin{eqnarray}\label{IDBianchi0}
\nabla_{0\mu} \star F_{0 \nu\rho}+\nabla_{0\nu} \star F_{0 \rho\mu}+\nabla_{0\rho} \star F_{0 \mu\nu}=0 \; ,
\end{eqnarray}
which completes the equations for the NC electromagnetism. The field equations for the $B_{0\mu\nu}$ tensor fields
are auxiliary equations that emerge exclusively from the NC $\theta$ extra-dimensions
\begin{eqnarray}\label{EqfieldB0}
\nabla_{0\mu\nu} \, \star \, G_{0}^{\;\;\mu\nu\rho\sigma}
-\frac{1}{2} \left(\nabla_{\mu}^{\;\; \,\rho} \star F_{0}^{\;\;\mu\sigma}
-\nabla_{\mu}^{\;\; \, \sigma} \star F_{0}^{\;\;\mu\rho}
\right)=\frac{1}{2} \; g^{\prime} \,  {\cal J}_{0}^{\; \, \rho\sigma} \; ,
\end{eqnarray}
where the anti-symmetric covariant derivative $\nabla_{0\mu\nu}$ is defined by
\begin{eqnarray}
\label{DmunuG}
\nabla_{0\mu\nu}\star G_{0}^{\; \;\mu\nu\rho\sigma}:=\lambda \, \partial_{\mu\nu}G_{0}^{\;\;\mu\nu\rho\sigma}
-g^{\prime}\left[B_{0\mu\nu},G_{0}^{\;\;\mu\nu\rho\sigma}\right]_{\star} \; .
\end{eqnarray}
The current $J_{0}^{\;\mu}$ is the classical source of the fermion field, that is, $J_{0}^{\;\mu}=\bar{\psi} \gamma^{\mu} \star \psi$.
The anti-symmetric current ${\cal J}_{0}^{\;\mu\nu}$, as it was showed in the earlier section, is the source for the classical
fermion field attached to the generator spin, ${\cal J}_{0}^{\;\mu\nu}=\bar{\psi}\Gamma^{\mu\nu}\star \psi$. This fact is due to the
extra-dimension of the NC space-time. When the NC parameter goes to zero, the usual current $J_{0}^{\;\mu}$
for the QED is recovered, while the
new current ${\cal J}_{0}^{\;\mu\nu}$ is automatically zero. From the Eqs. (\ref{EqfieldA0}) and (\ref{EqfieldB0}),
we can obtain the conservation law
\begin{eqnarray}\label{EqContinuidade0}
\nabla_{0\mu} \star J_{0}^{\; \mu}+\frac{1}{2} \; \nabla_{0\mu\nu}\star{\cal J}_{0}^{\;\;\mu\nu}=0 \; ,
\end{eqnarray}
which expresses the electric charge covariant conservation.

Analogously, the sector $SU^{\star}(N)$ from the Lagrangian (\ref{LSpinor+gauge}), gives us the NC Yang-Mills field
\begin{eqnarray}\label{EqfieldAYM}
\nabla_{\mu} \, \star \, F^{\mu\nu a}+\nabla_{\mu} \, \star \, G^{\mu\rho\nu \hspace{0.2cm} a}_{\hspace{0.45cm}\rho} =g \, J^{\nu a} \; ,
\end{eqnarray}
where the covariant derivative here $\nabla_{\mu}$ differ from $\nabla_{0\mu}$ by the commutator
in the adjoint representation
\begin{eqnarray}\label{DmuFmunu}
\nabla_{\mu} \star F^{\mu\nu a}:=\partial_{\mu}F^{\mu\nu a}-g\left[A_{\mu},F^{\mu\nu}\right]^{a}_{\star} \; ,
\end{eqnarray}
where this commutator is defined by
\begin{eqnarray}\label{commutadorAA}
\left. \left[A_{\mu},F^{\mu\nu}\right]^{a}_{\star}:= f^{abc} \,
\cos\left(\frac{\theta^{\alpha\beta}}{2}\,\partial_{\alpha}\partial_{\beta}^{\prime} \right)
A_{\mu}^{b}(x,\theta) F^{\mu\nu c}(x^{\prime},\theta) \right|_{x^{\prime}=x}
\nonumber \\
\left.
+ \, d^{abc} \, \sin\left(\frac{\theta^{\alpha\beta}}{2}\,\partial_{\alpha}\partial_{\beta}^{\prime} \right)
A_{\mu}^{b}(x,\theta) F^{\mu\nu c}(x^{\prime},\theta) \right|_{x^{\prime}=x} \; .
\end{eqnarray}
It is not difficult to see that the Bianchi identity is valid too, namely,
\begin{eqnarray}
\nabla_{\mu} \star F_{\nu\rho}^{\;\; \; a}+\nabla_{\nu} \star F_{\rho\mu}^{\; \; \; a}+\nabla_{\rho} \star F_{\mu\nu}^{\; \; \; a}=0 \; .
\end{eqnarray}
The field equation of $B^{\mu\nu a}$ is given by
\begin{eqnarray}\label{EqfieldsAB}
\nabla_{\mu\nu} \, \star \, G^{\mu\nu\rho\sigma a}
-\frac{1}{2} \left(\nabla_{\mu}^{\;\;\rho} \star F^{\mu\sigma a}-\nabla_{\mu}^{\;\;\sigma} \star F^{\mu\rho a}
\right)=\frac{1}{2} \; g^{\prime}{\cal J}^{\rho\sigma a} \; ,
\end{eqnarray}
where $\nabla_{\mu\nu}$ is the covariant derivative
\begin{eqnarray}
\label{DmunuG2}
\nabla_{\mu\nu}\star G^{\mu\nu\rho\sigma a}:=\lambda \, \partial_{\mu\nu}G^{\mu\nu\rho\sigma a}
-g^{\prime}\left[B_{\mu\nu},G^{\mu\nu\rho\sigma}\right]^{a}_{\star} \; ,
\end{eqnarray}
and this commutator is similar to expression (\ref{commutadorAA}). The non-Abelian currents of the previous equations are
 \begin{eqnarray}\label{Jmu}
J^{\mu a}(x,\theta)=\bar{\psi}_{i}(x,\theta)\gamma^{\mu} \, \left(t^{a}\right)_{ij}\star\psi_{j}
(x,\theta) \; ,
\end{eqnarray}
and the anti-symmetric current is
\begin{eqnarray}\label{Jmunu}
{\cal J}^{\mu\nu a}(x,\theta)=\bar{\psi}_{i}(x,\theta)\Gamma^{\mu\nu}\, \left(t^{a}\right)_{ij} \star \psi_{j}(x,\theta) \; ,
\end{eqnarray}
which obeys the continuity equation
\begin{eqnarray}\label{EqContinuidade}
\nabla_{\mu} \star J^{\mu a}+\frac{1}{2} \; \nabla_{\mu\nu}\star{\cal J}^{\mu\nu a}=0 \; .
\end{eqnarray}
%
%
%
%
%
%

Hence, we have obtained, separately, the field equations of the NC subgroups $U^{\star}(1)$ and $SU^{\star}(N)$ of $U^{\star}(N)$.
The DFR Yang-Mills field equations bring  a new result in DFR literature, and  we can reobtain the DFR electromagnetism classical field equations.
In this case, we can make $f^{abc}=0$ and $d^{abc}=1$ in the non-Abelian equations of $SU^{\star}(N)$ to obtain the correspondents equations of the subgroup $U^{\star}(1)$.
\section{Propagators in the DFR space-time}
In this section we will discuss  the quantum aspects of the new (in DFR formalism) Lagrangian in Eq. (\ref{LSpinor+gauge})
in order to compute the propagators of both gauge and fermion fields. The quantization of the model
has made us to introduce gauge fixing terms and ghost fields $\eta^{a}$ associated with
the NC non-Abelian gauge field in (\ref{LSpinor+gauge})
\begin{eqnarray}\label{LUNGhosts}
&&{\cal L}_{U^{\star}(N)}+{\cal L}_{gf}+ {\cal L}_{ghosts}=\bar{\psi}_{i} \, \star \, \Big(\,i\gamma^{\mu}D_{\mu}\star
+\,\frac{i}{2}\,\Gamma^{\mu\nu}D_{\mu\nu}\star-m \Big)_{ij} \psi_{j}
\nonumber \\
&&-\frac{1}{4}\,F_{0\mu\nu} \, \star \, F_{0}^{ \; \mu\nu}
-\frac{1}{2\xi_{0}} \, \left(\partial_{\mu}A_{0}^{\; \mu}\right)\star\left(\partial_{\nu}A_{0}^{\; \nu}\right)
\nonumber \\
&&-\frac{1}{4}\, G_{0\mu\nu\rho\sigma} \, \star \, G_{0}^{\; \; \mu\nu\rho\sigma}
-\frac{1}{2\alpha_{0}}\left(\partial_{\mu\nu}B_{0}^{ \; \; \mu\nu}\right)\star\left(\partial_{\rho\sigma}B_{0}^{ \; \; \rho\sigma}\right)
-\frac{1}{2}\,F_{0\mu\nu} \, \star \, G^{ \hspace{0.15cm} \mu\rho\nu}_{0\hspace{0.44cm}\rho}
\nonumber \\
&&\,-\frac{1}{4}\,F_{\mu\nu}^{\hspace{0.3cm} a} \, \star \, F^{\mu\nu a}
-\frac{1}{2\xi} \, \left(\partial_{\mu}A^{\mu a}\right) \star \left(\partial_{\nu}A^{\nu a}\right)
\nonumber \\
&&-\frac{1}{4}\, G_{\mu\nu\rho\sigma}^{\hspace{0.68cm} a}\, \star \, G^{\mu\nu\rho\sigma a}
-\frac{1}{2\alpha}\left(\partial_{\mu\nu}B^{\mu\nu a}\right)\star\left(\partial_{\rho\sigma}B^{\rho\sigma a}\right)
-\frac{1}{2}\,F_{\mu\nu}^{\hspace{0.3cm} a} \, \star \, G^{\mu\rho\nu \hspace{0.2cm} a}_{\hspace{0.45cm}\rho}
\nonumber \\
&&+\bar{\eta}^{\,a} \star \partial_{\mu}D^{\mu}\eta^{\,a}
+\bar{\eta}^{\,a} \star \frac{\lambda}{2} \, \partial_{\mu\nu}D^{\mu\nu}\eta^{\,a} \; ,
\end{eqnarray}
where $(\xi_{0}, \alpha_{0}, \xi, \alpha)$ are real parameters.
The ghost fields $\eta=\eta^{a}t^{a}$ and $\bar{\eta}=\bar{\eta}^{a}t^{a}$ have the local transformation
in the adjoint representation of the star gauge group given by
\begin{eqnarray}
\eta \longmapsto \eta^{\prime}=U(x,\theta) \star \eta \star U^{\dagger}(x,\theta)
\hspace{0.5cm} \mbox{and} \hspace{0.5cm}
\bar{\eta} \longmapsto \bar{\eta}^{\prime}=U(x,\theta) \star \bar{\eta} \star U^{\dagger}(x,\theta) \; .
\end{eqnarray}
The covariant derivatives that act on the ghost field are defined by
\begin{eqnarray}
D_{\mu}\eta^{a}=\partial_{\mu}\eta^{a}-g\left[A_{\mu},\eta\right]^{a}_{\star} \; \quad \mbox{and} \quad \;
D_{\mu\nu}\eta^{a}=\lambda\, \partial_{\mu\nu}\eta^{a}-g^{\prime}\left[B_{\mu\nu},\eta\right]^{a}_{\star} \; ,
\end{eqnarray}
where these commutator is like the ones in (\ref{commutadorAA}). The quantum action of the model is defined by the
integration of the Lagrangian (\ref{LUNGhosts}) through the volume of the $(x+\theta)$-space-time
%
%
\begin{eqnarray}\label{SQ}
S_{Quantum}=\int d^{4}x \, \, d^{6}\theta \, \, W(\theta) \left({\cal L}_{U^{\star}(N)}+{\cal L}_{gf}+ {\cal L}_{ghosts} \right) \; ,
\end{eqnarray}
%
%
which can be written as the sum of the free part with the interaction terms, that is, $S_{Quantum}=S_{0}+S_{int}$,
where the free part can be simplified using the identity (\ref{Idmoyalproduct2}). Therefore the free part of the action given by (\ref{SQ}) is
\begin{eqnarray}\label{SQEDDFRfree}
&&S_{0}=\int d^{4}x \, \, d^{6}\theta \, \, W(\theta) \, \left[ \, \bar{\psi}_{i}\,
\Big(\,i\gamma^{\mu}\partial_{\mu}+\,{{i\lambda}\over2}\,\Gamma^{\mu\nu}\partial_{\mu\nu}-m \Big)_{ij} \psi_{j}
\right. \nonumber \\
&&
\left.
-\frac{1}{4}\,\left(\partial_{\mu}A_{0\nu}-\partial_{\nu}A_{0\mu}\right)^{2}-\frac{1}{2\xi_{0}}\,\left(\partial_{\mu}A_{0}^{\; \,\mu}\right)^{2}
-\frac{\lambda^{2}}{4} \, \left(\partial_{\mu\nu}B_{0\rho\sigma}-\partial_{\rho\sigma}B_{0\mu\nu}\right)^{2}
-\frac{\lambda^{2}}{2\alpha_{0}} \, \left(\partial_{\mu\nu}B_{0}^{\; \, \mu\nu}\right)^{2}
\right. \nonumber \\
&&
\left.
-\frac{\lambda}{2} \, \left(\partial_{\mu\nu}B_{0}^{\; \, \rho\nu}-\partial^{\rho\nu}B_{0\mu\nu}\right)
\left(\partial^{\mu}A_{0\rho}-\partial_{\rho}A_{0}^{\; \,\mu} \right)
\right. \nonumber \\
&&
\left.
-\frac{1}{4}\,\left(\partial_{\mu}A^{a}_{\nu}-\partial_{\nu}A^{a}_{\mu}\right)^{2}-\frac{1}{2\xi}\,\left(\partial_{\mu}A^{\mu a}\right)^{2}
-\frac{\lambda^{2}}{4} \, \left(\partial_{\mu\nu}B^{a}_{\rho\sigma}-\partial_{\rho\sigma}B^{a}_{\mu\nu}\right)^{2}
-\frac{\lambda^{2}}{2\alpha} \, \left(\partial_{\mu\nu}B^{\mu\nu a}\right)^{2}
\right. \nonumber \\
&&
\left.
-\frac{\lambda}{2} \, \left(\partial_{\mu\nu}B^{\rho\nu a}-\partial^{\rho\nu}B_{\mu\nu}^{a}\right)
\left(\partial^{\mu}A^{a}_{\rho}-\partial_{\rho}A^{\mu a} \right)
+\bar{\eta}^{a} \left( \Box+\frac{\lambda^{2}}{2} \,\Box_{\theta} \right)\eta^{a}\right] \; .
\end{eqnarray}
To obtain the propagators, it is convenient to write the previous action in momentum space.
Using the Fourier integrals for both gauge and fermions fields, analogous to the scalar case in (\ref{phiXtheta}),
we have the action $S_{0}$ that can be written as
\begin{eqnarray}\label{S0IntFourier}
&&S_{0}=\int \frac{d^{4}p}{(2\pi)^{4}} \frac{d^{6}k}{(2\pi\lambda^{-2})^{6}}
\frac{d^{6}k^{\prime}}{(2\pi\lambda^{-2})^{6}} \; \; e^{-\frac{\lambda^{4}}{4}(k+k^{\prime})^{2}}
\times
\nonumber \\
&& \times \; \left\{ \bar{u}(p,k)\Big(\,\gamma^{\mu}p_{\mu}
+\,{{\lambda}\over2}\,\Gamma^{\mu\nu}k_{\mu\nu}^{\prime}
-m \Big) u(p,k^{\prime})
\right.
\nonumber \\
&&
\left.
-\bar{v}^{\, a}(p,k) \left[ p^{2}+\frac{\lambda^{2}}{2} \left( k\cdot k^{\prime} \right) \right] v^{\, a}(p,k^{\prime})
\right.
\nonumber \\
&&
\left.
+\frac{1}{2}\,a_{0}^{\; \,\mu}(p,k)\left[\eta_{\mu\nu}\,p^{2}+\left(\frac{1}{\xi_{0}}-1\right)p_{\mu}\,p_{\nu}\right]a_{0}^{\; \,\nu}(p,k^{\prime})
\right.
\nonumber \\
&&
\left.
+\frac{\lambda^{2}}{2} \, b_{0}^{\;\,\rho\lambda}(p,k)\left[ {\uma_{\rho\lambda\sigma\kappa}} \left(k\cdot k^{\prime}\right)
+\left(\frac{1}{\alpha_{0}}-1\right) k_{\rho\lambda} \, k_{\sigma\kappa}^{\prime}\right] b_{0}^{\;\,\sigma\kappa}(p,k^{\prime})
\right.
\nonumber \\
&&
\left.
+\frac{\lambda}{4} \, a_{0}^{\; \,\mu}(p,k) \, p^{\alpha} \left(\eta_{\mu\sigma} k^{\prime}_{\alpha\kappa}-\eta_{\mu\kappa} k^{\prime}_{\alpha\sigma}
-\eta_{\alpha\sigma}  k^{\prime}_{\mu\kappa}+\eta_{\alpha\kappa}  k^{\prime}_{\mu\sigma} \right) b_{0}^{\; \, \sigma\kappa}(p,k^{\prime})
\right.
\nonumber \\
&&
\left.
+\frac{\lambda}{4} \, b_{0}^{\;\,\rho\lambda}(p,k)
\, p^{\alpha} \, \left( \eta_{\nu\rho} k_{\alpha\lambda}-\eta_{\nu\lambda} k_{\alpha\rho}
-\eta_{\alpha\rho} k_{\nu\lambda}+\eta_{\alpha\lambda} k_{\nu\rho} \right) a_{0}^{\;\,\nu}(p,k^{\prime})
\right.
\nonumber \\
&&
\left.
+\frac{1}{2}\,a^{\mu a}(p,k)\left[\eta_{\mu\nu}\,p^{2}+\left(\frac{1}{\xi}-1\right)p_{\mu}\,p_{\nu}\right]a^{\nu a}(p,k^{\prime})
\right.
\nonumber \\
&&
\left.
+\frac{\lambda^{2}}{2} \, b^{\rho\lambda a}(p,k)\left[ {\uma_{\rho\lambda\sigma\kappa}} \left(k\cdot k^{\prime}\right)
+\left(\frac{1}{\alpha}-1\right) k_{\rho\lambda} \, k_{\sigma\kappa}^{\prime}\right] b^{\sigma\kappa a}(p,k^{\prime})
\right.
\nonumber \\
&&
\left.
+\frac{\lambda}{4} \, a^{\mu a}(p,k) \, p^{\alpha} \left(\eta_{\mu\sigma} k^{\prime}_{\alpha\kappa}-\eta_{\mu\kappa} k^{\prime}_{\alpha\sigma}
-\eta_{\alpha\sigma}  k^{\prime}_{\mu\kappa}+\eta_{\alpha\kappa}  k^{\prime}_{\mu\sigma} \right) b^{\sigma\kappa a}(p,k^{\prime})
\right.
\nonumber \\
&&
\left.
+\frac{\lambda}{4} \, b^{\rho\lambda a}(p,k)
\, p^{\alpha} \, \left( \eta_{\nu\rho} k_{\alpha\lambda}-\eta_{\nu\lambda} k_{\alpha\rho}
-\eta_{\alpha\rho} k_{\nu\lambda}+\eta_{\alpha\lambda} k_{\nu\rho} \right) a^{\nu a}(p,k^{\prime}) \right\} \; ,
\end{eqnarray}
where $\uma_{\rho\lambda\sigma\kappa}$
is the anti-symmetrized identity matrix previously defined in the first section, and $u$, $v$, $a_{0}^{\mu}$, $a^{\mu a}$,
$b_{0}^{\mu\nu}$, $b^{\mu\nu a}$ are Fourier transforms of the fermions, ghosts and gauge fields, respectively.
For both fermionic and ghost sectors, the propagators are given by the inverse of their kinetic terms, so we have that
\begin{eqnarray}\label{PropagatorPsi}
\langle \bar{u}(p,k) \; u(p,k^{\prime}) \rangle \! &=& \! (2\pi)^{4}\delta^{(4)}(p-p^{\prime}) \,\, \frac{i}
{\gamma^{\mu}p_{\mu}+\frac{\lambda}{2} \,\Gamma^{\mu\nu}k_{\mu\nu}^{\,\prime}-m} \; ,
\nonumber \\
\langle \bar{v}^{\, a}(p,k) \; v^{\, b}(p,k^{\prime}) \rangle \! &=& \! \frac{-i \, \delta^{ab}}{p^{2}+\frac{\lambda^{2}}{2} \left( k\cdot k^{\prime} \right)} \; .
\end{eqnarray}
In the gauge fields sector, we consider $\left(a_{0}^{\; \,\mu}, b_{0}^{\; \, \mu \nu}\right)$ and
$\left(a^{\mu a}, b^{\mu\nu a}\right)$ as independent fields, so that the result of the propagators
for NC non-Abelian gauge fields is similar to the NC Abelian one. We can write the Abelian part
$\left(a_{0}^{\; \,\mu}, b_{0}^{\; \, \mu \nu}\right)$ of (\ref{S0IntFourier}) in the matrix form
\begin{equation}\label{MatrizAB}
\frac{1}{2}\left(\begin{array}{cc}
a_{0}^{\;\,\mu} & b_{0}^{\; \,\rho\lambda} \end{array}\right)\!\left(\begin{array}{cc}
p^{2}\left(P_{\mu\nu}+ \xi^{-1} T_{\mu\nu} \right) & \,Q_{\mu\sigma\kappa}(p,k^{\prime}) \\
\,Q_{\nu\rho\lambda}(p,k) & \lambda^{2} \left(k\cdot k^{\prime} \right)\left(M_{\rho\lambda\sigma\kappa}
+\alpha^{-1}N_{\rho\lambda\sigma\kappa} \right) \end{array}\right)
\!\left(\begin{array}{c}
a_{0}^{\; \,\nu}\\
b_{0}^{\; \,\sigma\kappa}
\end{array}\right) \; ,
\end{equation}
and the projectors $P$, $T$, $M$ and $N$ are defined by
\begin{eqnarray}\label{PTQ}
&&P_{\mu\nu}(p)=\eta_{\mu\nu}-\frac{p_{\mu}p_{\nu}}{p^{2}}
\hspace{0.2cm} , \hspace{0.2cm}
T_{\mu\nu}(p)= \frac{p_{\mu}p_{\nu}}{p^{2}} \;\; ,
\nonumber \\
&&M_{\mu\nu\rho\lambda}(k,k^{\prime})={\uma_{\mu\nu\rho\lambda}}-\frac{k_{\mu\nu}k_{\rho\lambda}^{\prime}}{k\cdot k^{\prime}}
\hspace{0.2cm} , \hspace{0.2cm}
N_{\mu\nu\rho\lambda}(k,k^{\prime})=\frac{k_{\mu\nu}k_{\rho\lambda}^{\prime}}{k\cdot k^{\prime}} \;\; ,
\end{eqnarray}
and the matrix element $Q$ is
\begin{eqnarray}\label{Q}
Q_{\mu\sigma\kappa}(p,k)= \frac{\lambda}{2} \, p^{\alpha}
\left( \eta_{\mu\sigma} k_{\alpha\kappa}-\eta_{\mu\kappa} k_{\alpha\sigma}
- \eta_{\alpha\sigma}  k_{\mu\kappa}
+ \eta_{\alpha\kappa}  k_{\mu\sigma} \right) \; ,
\end{eqnarray}
where we can observe that it is anti-symmetrized in the index $(\sigma, \kappa)$. The contraction $k_{\mu\nu}k^{\mu\nu \prime}$
has been simplified by the scalar product $k\cdot k^{\prime}$ using a short notation. It is easy to show that the projectors satisfy
the relations
\begin{eqnarray}\label{RelProj}
&&P_{\mu\nu}P^{\mu\rho}=P_{\nu}^{\;\;\rho}
\hspace{0.2cm} , \hspace{0.2cm}
T_{\mu\nu}T^{\mu\rho}=T_{\nu}^{\;\;\rho}
\hspace{0.2cm} , \hspace{0.2cm}
M_{\mu\nu\alpha\beta} M^{\alpha\beta}_{\;\;\;\;\;\;\rho\lambda}=M_{\mu\nu\rho\lambda} \; ,
\hspace{1.5cm}
\nonumber \\
&&N_{\mu\nu\alpha\beta}\, N^{\alpha\beta}_{\;\;\;\;\;\;\rho\lambda}=N_{\mu\nu\rho\lambda}
\; , \hspace{0.2cm}
P_{\mu\sigma}\,Q^{\sigma}_{\;\;\; \nu\rho}=Q_{\mu\nu\rho}
\hspace{0.2cm} , \hspace{0.2cm}
Q^{\mu\alpha\beta} \, M_{\alpha\beta\nu\rho}=Q^{\mu}_{\;\;\;\nu\rho}
\; , \hspace{0.8cm}
\nonumber \\
&&P_{\mu\nu}\,T^{\nu\rho}=M_{\mu\nu\alpha\beta} N^{\alpha\beta}_{\hspace{0.38cm}\rho\lambda}
=P_{\mu\nu}N^{\mu\nu}_{\hspace{0.35cm}\rho\sigma}=
T_{\mu\nu}N^{\mu\nu}_{\hspace{0.35cm}\rho\sigma}=
T_{\mu\sigma}Q^{\sigma}_{\hspace{0.23cm}\nu\rho}=
Q^{\mu\alpha\beta}N_{\alpha\beta\nu\rho}=0 \; ,
\hspace{1cm}
\end{eqnarray}
and the possible contractions that comprises $Q$ will be given by
\begin{eqnarray}
Q_{\mu\sigma\kappa}(p,k^{\prime})Q_{\nu}^{\;\;\,\sigma\kappa}(p,k)
&=&\frac{\lambda^{2}}{2} \, p^{2} \left(k \cdot k^{\prime} \right) \, N_{\mu\alpha\nu\beta}\left(P^{\alpha\beta}-T^{\alpha\beta}\right) \; ,
\nonumber \\
Q_{\alpha\mu\nu}(p,k)Q^{\alpha}_{\;\;\,\rho\sigma}(p,k^{\prime})&=&\frac{\lambda^{2}}{4}\, p^{2} \left(k \cdot k^{\prime}\right)
\left[\phantom{\frac{1}{2}}\!\!\!\eta_{\mu\rho}T^{\alpha\beta}N_{\alpha\nu\beta\sigma}-\eta_{\mu\sigma}T^{\alpha\beta}N_{\alpha\nu\beta\rho}
\right. \nonumber \\
&&
\left.
-\eta_{\nu\rho}T^{\alpha\beta}N_{\alpha\mu\beta\sigma}
+\eta_{\nu\sigma}T^{\alpha\beta}N_{\alpha\mu\beta\rho}
+ T_{\mu}^{\;\;\alpha}N_{\sigma\nu\alpha\rho}
\right. \nonumber \\
&&
\left.
-T_{\nu}^{\;\;\alpha}N_{\sigma\mu\alpha\rho}
-T_{\mu}^{\;\;\alpha}N_{\rho\nu\alpha\sigma}
+T_{\nu}^{\;\;\alpha}N_{\rho\mu\alpha\sigma}
+T_{\sigma}^{\;\;\alpha}N_{\alpha\nu\mu\rho}
\right. \nonumber \\
&&
\left.
-T_{\sigma}^{\;\;\alpha}N_{\alpha\mu\nu\rho}-T_{\rho}^{\;\;\alpha}N_{\alpha\nu\mu\sigma}+T_{\rho}^{\;\;\alpha}N_{\alpha\mu\nu\sigma}
+T_{\mu\rho}N_{\alpha\nu\;\;\,\sigma}^{\;\;\;\;\alpha}
\right. \nonumber \\
&&
\left.
-T_{\mu\sigma}N_{\alpha\nu\;\;\,\rho}^{\;\;\;\;\alpha}
-T_{\nu\rho}N_{\alpha\mu\;\;\,\sigma}^{\;\;\;\;\alpha}
+T_{\nu\sigma}N_{\alpha\mu\;\;\,\rho}^{\;\;\;\;\alpha}
\phantom{\frac{1}{2}}\!\!\!\right] \; .
\end{eqnarray}
These previous properties permit us to invert the matrix $2 \times 2$ in (\ref{MatrizAB}) to give us the
Abelian propagators fields $A_{0}^{\, \mu}$ and $B_{0}^{\; \mu\nu}$
\begin{eqnarray}\label{PropagadoresAB0}
\langle a_{0\mu}(p,k) \; a_{0\nu}(p,k^{\prime}) \rangle &=&-\frac{i \, \delta^{ab}}{p^{2}}\left[\eta_{\mu\nu}
+\left(\xi_{0}-1\right)\frac{p_{\mu}p_{\nu}}{p^{2}} \right]
+\frac{2i \, \delta_{ab}}{p^{2}} \left(\eta^{\alpha\beta}-2\,\frac{p^{\alpha}p^{\beta}}{p^{2}} \right) \frac{k_{\mu\alpha} k^{\prime}_{\nu\beta}}{k\cdot k^{\prime}}
\nonumber \\
\langle b_{0\mu\nu}(p,k) \; b_{0\rho\sigma}(p,k^{\prime}) \rangle &=& -\frac{i \, \delta^{ab}}{\lambda^{2}(k\cdot k^{\prime})}
\left[{\uma_{\mu\nu\rho\lambda}}+\left(\alpha_{0}-1\right)\,\frac{k_{\mu\nu}k_{\rho\lambda}^{\prime}}{k\cdot k^{\prime}}
-4 \, \mathbb{K}_{\mu\nu\rho\sigma}(p,k,k^{\prime}) \right] \, ,
\hspace{1cm}
\end{eqnarray}
where
\begin{eqnarray}\label{K}
&&\mathbb{K}_{\mu\nu\rho\sigma}(p,k,k^{\prime}):=\eta_{\mu\rho}T^{\alpha\beta}N_{\alpha\nu\beta\sigma}
-\eta_{\mu\sigma}T^{\alpha\beta}N_{\alpha\nu\beta\rho}
-\eta_{\nu\rho}T^{\alpha\beta}N_{\alpha\mu\beta\sigma}
\nonumber \\
&&+\eta_{\nu\sigma}T^{\alpha\beta}N_{\alpha\mu\beta\rho}
+ T_{\mu}^{\;\;\alpha}N_{\sigma\nu\alpha\rho}-T_{\nu}^{\;\;\alpha}N_{\sigma\mu\alpha\rho}
-T_{\mu}^{\;\;\alpha}N_{\rho\nu\alpha\sigma}
\nonumber \\
&&+T_{\nu}^{\;\;\alpha}N_{\alpha\nu\mu\sigma}
+T_{\sigma}^{\;\;\alpha}N_{\alpha\nu\mu\rho}
-T_{\sigma}^{\;\;\alpha}N_{\alpha\mu\nu\rho}
-T_{\rho}^{\;\;\alpha}N_{\alpha\nu\mu\sigma}
\nonumber \\
&&+T_{\rho}^{\;\;\alpha}N_{\alpha\mu\nu\sigma}
+T_{\mu\rho}N_{\alpha\nu\;\;\,\sigma}^{\;\;\;\;\alpha}
-T_{\mu\sigma}N_{\alpha\nu\;\;\,\rho}^{\;\;\;\;\alpha}
-T_{\nu\rho}N_{\alpha\mu\;\;\,\sigma}^{\;\;\;\;\alpha}
+T_{\nu\sigma}N_{\alpha\mu\;\;\,\rho}^{\;\;\;\;\alpha} \; \; \; ,
\end{eqnarray}
and the mixed propagator is
\begin{eqnarray}\label{MixPropagadoresAB}
\langle a_{0\mu}(p,k) \; b_{0\sigma\kappa}(p,k^{\prime}) \rangle=
\langle b_{0\sigma\kappa}(p,k) \; a_{0\mu}(p,k^{\prime}) \rangle = \frac{i}{p^{2}}
\, \frac{Q_{\mu\sigma\kappa}(p,k)}{ \lambda^{2} \,(k\cdot k^{\prime})} \; .
\end{eqnarray}
For the non-Abelian gauge fields, we can obtain the propagators multiplying the previous one by $\delta^{ab}$
\begin{eqnarray}\label{PropagadoresAB}
\langle a_{\mu}^{\; a}(p,k) \; a_{\nu}^{\; b}(p,k^{\prime}) \rangle &=&-\frac{i \, \delta^{ab}}{p^{2}}
\left[\eta_{\mu\nu}+\left(\xi-1\right)\frac{p_{\mu}p_{\nu}}{p^{2}} \right]
+\frac{2i \, \delta_{ab}}{p^{2}} \left(\eta^{\alpha\beta}-2\,\frac{p^{\alpha}p^{\beta}}{p^{2}} \right) \frac{k_{\mu\alpha} k^{\prime}_{\nu\beta}}{k\cdot k^{\prime}}
\nonumber \\
\langle b_{\mu\nu}^{\; \, a}(p,k) \; b_{\rho\sigma}^{\; \, b}(p,k^{\prime}) \rangle &=& -\frac{i \, \delta^{ab}}{\lambda^{2}(k\cdot k^{\prime})}
\left[{\uma_{\mu\nu\rho\lambda}}+(\alpha-1)\,\frac{k_{\mu\nu}k_{\rho\lambda}^{\prime}}{k\cdot k^{\prime}}
-4 \, \mathbb{K}_{\mu\nu\rho\sigma}(p,k,k^{\prime}) \right] \; , \; \; \;
\nonumber \\
\langle a_{\mu}^{\; a}(p,k) \; b_{\sigma\kappa}^{\; \, b}(p,k^{\prime}) \rangle &=&
\langle b_{\sigma\kappa}^{\; a}(p,k) \; a_{\mu}^{\; b}(p,k^{\prime}) \rangle= \frac{i\, \delta^{ab}}{p^{2}}
\frac{Q_{\mu\sigma\kappa}(p,k)}{ \lambda^{2} \,\left(k \cdot k^{\prime}\right)} \; .
\end{eqnarray}

The interaction terms $S_{int}$ in (\ref{SQ}) are given by
\begin{eqnarray}\label{SQEDDFR}
&&S_{int}=
\int d^{4}x \, \, d^{6}\theta \, \, W(\theta) \left\{\phantom{\frac{1}{4}}
\!\!\!\!\!\!  -g \, \bar{\psi}\star \gamma^{\mu}A_{0\mu} \star\psi + g \, \partial_{\mu}A_{0\nu} \,
\left[A_{0}^{\;\mu},A_{0}^{\;\nu}\right]_{\star}
\!-\frac{1}{4} \, g^{2} \, \left[A_{0\mu},A_{0\nu}\right]_{\star}^{\;2}
\right.
\nonumber \\
&&
\left.
-g^{\prime} \, \bar{\psi}\star\Gamma^{\mu\nu}B_{0\mu\nu}\star\psi+g^{\prime} \, \partial_{\mu\nu}B_{0\rho\sigma} \, \left[B_{0}^{\; \mu\nu},B_{0}^{\; \rho\sigma}\right]_{\star}
-\frac{1}{4} \, g^{\prime\,2} \left[B_{0\mu\nu} , B_{0\rho\sigma} \right]_{\star}^{\; 2}
\right.
\nonumber \\
&&
\left.
-g \, \partial_{\mu\nu}B_{0}^{\; \, \rho\nu} \, \left[A_{0}^{\; \mu},A_{0\rho}\right]_{\star}
-g^{\prime} \, \partial_{\mu}A_{0}^{\;\rho} \, \left[B_{0}^{\; \, \mu\nu},B_{0\rho\nu}\right]_{\star}
+\frac{1}{2} \, g \, g^{\prime} \, \left[A_{0}^{\mu},A_{0\rho}\right]_{\star} \, \left[B_{0\mu\nu} , B_{0}^{\; \,\rho\nu} \right]_{\star}
\right.
\nonumber \\
&&
\left.
+g\, \partial_{\mu}\bar{\eta}^{a}\left[A_{0}^{\; \mu},\eta\right]_{\star}
+\frac{1}{2} \, g^{\prime}\, \partial_{\mu\nu}\bar{\eta}^{a}\left[B_{0}^{\; \mu\nu},\eta\right]_{\star}
\right.
\nonumber \\
&&
\left.
-g \, \bar{\psi}\star \gamma^{\mu}A_{\mu}^{\;\;a}t^{a}\star\psi+g \, \partial_{\mu}A^{a}_{\nu} \, \left[A^{\mu},A^{\nu}\right]^{a}_{\star}
-\frac{1}{4} \, g^{2} \, \left[A_{\mu},A_{\nu}\right]^{a}_{\star} \, \left[A^{\mu},A^{\nu}\right]^{a}_{\star}
\right.
\nonumber \\
&&
\left.
+g^{\prime} \, \partial_{\mu\nu}B_{\rho\sigma}^{a} \, \left[B^{\mu\nu},B^{\rho\sigma}\right]^{a}_{\star}
-\frac{1}{4} \, g^{\prime\,2} \left[B_{\mu\nu} , B_{\rho\sigma} \right]^{a}_{\star} \, \left[B^{\mu\nu} , B^{\rho\sigma} \right]^{a}_{\star}
\right.
\nonumber \\
&&
\left.
-g^{\prime} \, \bar{\psi}\star\Gamma^{\mu\nu}B_{\mu\nu}^{\;\;\; a}t^{a}\star\psi
+g\, \partial_{\mu}\bar{\eta}^{a}\left[A^{\mu},\eta\right]^{a}_{\star}
+\frac{1}{2} \, g^{\prime}\, \partial_{\mu\nu}\bar{\eta}^{a}\left[B^{\mu\nu},\eta\right]^{a}_{\star}
\right.
\nonumber \\
&&
\left.
-g \, \partial_{\mu\nu}B^{\rho\nu a} \, \left[A^{\mu},A_{\rho}\right]^{a}_{\star}
-g^{\prime} \, \partial_{\mu}A^{\rho a} \, \left[B^{\mu\nu},B_{\rho\nu}\right]^{a}_{\star}
+\frac{1}{2} \, g \, g^{\prime} \, \left[A^{\mu},A_{\rho}\right]^{a}_{\star} \, \left[B_{\mu\nu} , B^{\rho\nu} \right]^{a}_{\star} \,
\right\} \; . \; \; \;
\end{eqnarray}
After a tedious computation of the vertices rules in the momentum space for example, the $3$-line vertex of the gauge field $A^{\mu a}$ which is represented by
%
%
\begin{figure}[!h]
\begin{center}
\newpsobject{showgrid}{psgrid}{subgriddiv=1,griddots=10,gridlabels=6pt}
\begin{pspicture}(0,0.8)(4,3.2)
\psset{arrowsize=0.2 2}
\psset{unit=1}
%
%
%
\pscoil[coilarm=0,coilwidth=0.2,coilheight=1.0,linecolor=black](-1.5,1)(1.5,1)
\pscoil[coilarm=0,coilwidth=0.2,coilheight=1.0,linecolor=black](-0.1,1.1)(-0.1,3.1)
\put(0.2,2.9){\large$\partial_{\mu} A_{\nu}^{\; a}$}
\put(-1.5,1.3){\large$A^{\mu b}$}
\put(1.1,1.3){\large$A^{\nu c}$}
\put(2.2,1.2){\large$ g \, \partial_{\mu}A^{a}_{\nu} \, \left[A^{\mu},A^{\nu}\right]^{a}_{\star} \; .$}
\put(-1.2,2.9){$p_{1},k_{1}$}
\put(-1.8,0.5){$p_{2},k_{2}$}
\put(1,0.5){$p_{3},k_{3}$}
%
%
\end{pspicture}
%
%
%
\end{center}
\end{figure}

\noindent
This $3$-line vertex in the momentum space has the expression
\begin{eqnarray}
\label{Vabcmunurho}
&&V^{abc \, \mu\nu\rho}(p_{1},p_{2},p_{3};k_{1},k_{2},k_{3})=ig \, (2\pi)^{4} \delta^{(4)}(p_{1}+p_{2}+p_{3}) \,
e^{-\frac{\lambda^{4}}{4}\left(k_{1}+k_{2}+k_{3}\right)^{2}} \times
\nonumber \\
&&\times \left[ f^{abc} F^{\mu\nu\rho}(p_{1},p_{2},p_{3};k_{1},k_{2},k_{3})
+ i \, d^{abc} \, G^{\mu\nu\rho}(p_{1},p_{2},p_{3};k_{1},k_{2},k_{3}) \right] \; ,
\end{eqnarray}
where we can define the functions
\begin{eqnarray}
&&F^{\mu\nu\rho}\!:=\left(\eta^{\mu\rho} \, p_{1}^{\nu}-\eta^{\mu\nu} \, p_{1}^{\rho}\right) \, e^{-\lambda^{4}\left( \frac{p_{2}^{\mu}p_{3}^{\nu}-p_{2}^{\nu}p_{3}^{\mu}}{4} \right)^{2}}
\cosh\left[\frac{\lambda^{4}}{4}(k_{1}+k_{2}+k_{3})_{\mu\nu}(p_{2}^{\mu}p_{3}^{\nu}-p_{2}^{\nu}p_{3}^{\mu}) \right]
\nonumber \\
&&+\left(\eta^{\mu\nu} \, p_{2}^{\rho}-\eta^{\nu\rho} \, p_{2}^{\mu} \right)
e^{-\lambda^{4}\left( \frac{p_{1}^{\mu}p_{3}^{\nu}-p_{1}^{\nu}p_{3}^{\mu}}{4} \right)^{2}}
\!\!\!\cosh\left[\frac{\lambda^{4}}{4}(k_{1}+k_{2}+k_{3})_{\mu\nu}(p_{1}^{\mu}p_{3}^{\nu}-p_{1}^{\nu}p_{3}^{\mu}) \right]
\nonumber \\
&&+\left(\eta^{\nu\rho} \, p_{3}^{\mu}-\eta^{\mu\rho} \, p_{3}^{\nu}\right) \, e^{-\lambda^{4}\left( \frac{p_{1}^{\mu}p_{2}^{\nu}-p_{1}^{\nu}p_{2}^{\mu}}{4} \right)^{2}}
\!\!\cosh\left[\frac{\lambda^{4}}{4}(k_{1}+k_{2}+k_{3})_{\mu\nu}(p_{1}^{\mu}p_{2}^{\nu}-p_{1}^{\nu}p_{2}^{\mu})\right] \; ,
\hspace{0.5cm}
\end{eqnarray}
and
\begin{eqnarray}
&&G^{\mu\nu\rho}:=\left(\eta^{\mu\rho} \, p_{1}^{\nu}-\eta^{\mu\nu} \, p_{1}^{\rho} \right) \, e^{-\lambda^{4}\left( \frac{p_{2}^{\mu}p_{3}^{\nu}
-p_{2}^{\nu}p_{3}^{\mu}}{4} \right)^{2}}
\!\!\sinh\left[\frac{\lambda^{4}}{4}(k_{1}+k_{2}+k_{3})_{\mu\nu}(p_{2}^{\mu}p_{3}^{\nu}-p_{2}^{\nu}p_{3}^{\mu}) \right]
\nonumber \\
&&+\left(\eta^{\nu\rho} \, p_{2}^{\mu}-\eta^{\mu\nu} \, p_{2}^{\rho}\right) \,
e^{-\lambda^{4}\left( \frac{p_{1}^{\mu}p_{3}^{\nu}-p_{1}^{\nu}p_{3}^{\mu}}{4} \right)^{2}}
\!\!\sinh\left[\frac{\lambda^{4}}{4}(k_{1}+k_{2}+k_{3})_{\mu\nu}(p_{1}^{\mu}p_{3}^{\nu}-p_{1}^{\nu}p_{3}^{\mu}) \right]
\nonumber \\
&&+\left(\eta^{\nu\rho} \, p_{3}^{\mu}-\eta^{\mu\rho} \, p_{3}^{\nu} \right) \, e^{-\lambda^{4}\left( \frac{p_{1}^{\mu}p_{2}^{\nu}-p_{1}^{\nu}p_{2}^{\mu}}{4} \right)^{2}}
\!\!\!\sinh\left[\frac{\lambda^{4}}{4}(k_{1}+k_{2}+k_{3})_{\mu\nu}(p_{1}^{\mu}p_{2}^{\nu}-p_{1}^{\nu}p_{2}^{\mu})\right] \; .
\hspace{0.5cm}
\end{eqnarray}
The $3$-line vertex of the gauge field $A_{0}^{\; \mu}$ can be obtained by making $f^{abc}=0$ and $d^{abc}=1$ in (\ref{Vabcmunurho}),
so we have
%
%
\begin{figure}[!h]
\begin{center}
\newpsobject{showgrid}{psgrid}{subgriddiv=1,griddots=10,gridlabels=6pt}
\begin{pspicture}(0,0.8)(3.5,3)
\psset{arrowsize=0.2 2}
\psset{unit=1}
%
%
%
\pscoil[coilarm=0,coilaspect=0,coilwidth=0.2,coilheight=1.0,linecolor=black](-1.5,1)(1.5,1)
\pscoil[coilarm=0,coilaspect=0,coilwidth=0.2,coilheight=1.0,linecolor=black](-0.1,1.1)(-0.1,3.1)
\put(0.2,2.9){\large$\partial_{\mu} A_{0\nu}$}
\put(-1.5,1.3){\large$A_{0}^{\; \mu}$}
\put(1.1,1.3){\large$A_{0}^{\; \nu}$}
\put(2.2,1.2){\large$ g \, \partial_{\mu}A_{0\nu} \, \left[A_{0}^{\; \mu},A_{0}^{\; \nu}\right]_{\star} \; ,$}
\put(-1.2,2.9){$p_{1},k_{1}$}
\put(-1.8,0.5){$p_{2},k_{2}$}
\put(1,0.5){$p_{3},k_{3}$}
%
%
\end{pspicture}
%
%
%
\end{center}
\end{figure}

\noindent
where
\begin{eqnarray}\label{Vabcmunurho2}
V^{0 \, \mu\nu\rho}(p_{1},p_{2},p_{3};k_{1},k_{2},k_{3})=-g \, (2\pi)^{4} \delta^{(4)}(p_{1}+p_{2}+p_{3}) \, \times
\nonumber \\
\times \, e^{-\frac{\lambda^{4}}{4}\left(k_{1}+k_{2}+k_{3}\right)^{2}} G^{\mu\nu\rho}(p_{1},p_{2},p_{3};k_{1},k_{2},k_{3}) \; .
\end{eqnarray}
In the commutative limit, when $\lambda \rightarrow 0$, it can be verified that all the vertex of the gauge field $A^{\mu a}$ tends to the usual Yang-Mills case, and the self-interaction vertex of $A_{0}^{\; \mu}$ goes to zero. It is also interesting to realize that the vertex
of the gauge fields $B_{0\mu\nu}$ and $B_{\mu\nu}^{\; a}$ which interacts with the fermions $\psi_{i}$.
%
%
\begin{figure}[!h]
\begin{center}
\newpsobject{showgrid}{psgrid}{subgriddiv=1,griddots=10,gridlabels=6pt}
\begin{pspicture}(0,0.3)(3.5,3.2)
\psset{arrowsize=0.2 2}
\psset{unit=1}
%
%
\pscoil[coilarm=0,coilaspect=0,coilwidth=0.3,coilheight=1.0,linecolor=black](-1.5,1)(-1.5,3)
\psline[linecolor=black,linewidth=0.5mm]{-}(-3,1)(0,1)
\psline[linecolor=black,linewidth=0.5mm]{->}(-3,1)(-2,1)
\psline[linecolor=black,linewidth=0.5mm]{->}(-2,1)(-0.55,1)
\put(-1.2,2.8){\large$B_{0\mu\nu}$}
\put(-3,1.2){\large$\bar{\psi}_{i}$}
\put(-0.2,1.2){\large$\psi_{j}$}
\put(-4.4,0.2){\large$-g^{\prime} \, \bar{\psi}_{i}\star\Gamma^{\mu\nu}B_{0\mu\nu}\left(\delta_{ij}\right) \star \psi_{j}  $}
%
%
\pscoil[coilarm=0,coilwidth=0.2,coilheight=1.0,linecolor=black](4.5,1)(4.5,3)
\psline[linecolor=black,linewidth=0.5mm]{-}(3,1)(6,1)
\psline[linecolor=black,linewidth=0.5mm]{->}(3,1)(4,1)
\psline[linecolor=black,linewidth=0.5mm]{->}(5,1)(5.55,1)
\put(4.8,2.8){\large$B_{\mu\nu}^{\; a}$}
\put(2.9,1.2){\large$\bar{\psi}_{i}$}
\put(5.7,1.2){\large$\psi_{j}$}
\put(1.8,0.2){\large$-g^{\prime} \, \bar{\psi}_{i}\star\Gamma^{\mu\nu}B_{\mu\nu}^{\; \, a}\left(t^{a}\right)_{ij} \star \psi_{j} $}
\end{pspicture}
%
%
%
\end{center}
\end{figure}

\noindent
This is a new interaction due to both the noncommutativity, and the propagation in the $\theta$-space. The spin generator represented here by
$\Gamma^{\mu\nu}$ is coupled to an anti-symmetric gauge fields $B_{0\mu\nu}$ and $B_{\mu\nu}^{\; \, a}$.

\section{Conclusions and perspectives}
In this work we believe that some new steps were provided in order to fathom the DFR formalism which is considered in the NC literature as a possible path way to quantum gravity.
Keeping all these quantum ideas in mind, we have considered gauge Abelian and non-Abelian fields using the recent DFR framework where the parameter that carries the  noncommutativity feature, $\theta^{\mu\nu}$, represents independent degrees of freedom completing the DFR $D=10$ extended DFR space, which phase-space has the momentum $K$ associated with
$\theta$.

In this way we have started  with a first quantized formalism, where $\theta^{\mu\nu}$
and its canonical momentum $K_{\mu\nu}$ are operators  living in an extended Hilbert space.
This structure, which is compatible with the minimal canonical extension of the so-called DFR algebra,
is also invariant under an extended Poincar\'e group of symmetry, but keeping, among others, the usual Casimir invariant operators. After that, in a second quantized formalism scenario, we have succeed in presenting an explicit form for the extended Poincar\'e
generators and the same algebra of the first quantized description has been generated via generalized Heisenberg relations.  This is a basic point because the usual Casimir operators for the Poincar\'e group are proven to be kept, permitting to maintain the usual classification scheme for the elementary particles.  The next step in this program was to construct the mode expansion in order to represent the fields in terms of annihilation and creation operators, acting on some Fock space to be properly defined.

After that, in order to complete the DFR fermionic formalism given in \cite{Amorim2} we have constructed the Gamma matrices, its algebra and the DFR Dirac equation were also analyzed.

These results set the stage to discuss the gauge invariance subject in DFR scenario where star-covariant derivatives were used in order to construct the DFR gauge transformations. These ones allow us to construct DFR gauge invariant Lagrangians for the DFR versions of the QED and Yang-Mills models.   Nest, the Abelian and non-Abelian currents were calculated. We have seen that the Maxwell and Yang-Mills came out naturally from these DFR versions.   At the same time we have seen that, as in the standard commutative stage, the Abelian and non-Abelian models are connected, which confirm the correctness of the procedure.

The DFR NC conjecture has revealed the existence of an anti-symmetric tensor gauge field, beyond the NC
vector gauge field, to maintain the gauge invariance in the $\theta$-space. It is easy to see that in the commutative limit, any influence of this anti-symmetric field in the model goes to zero. A study of quantum aspects of this new field motivates our research in the future.

Finally, we have computed the respective propagators and the DFR point of view, diagrams, were we can view an analogy to the Feynman ones.

\section{Acknowledgments}

\noindent EMCA would like to thank CNPq (Conselho Nacional de Desenvolvimento Cient\'{\i}fico e Tecnol\'ogico), Brazilian research support federal agency, for partial financial support.

\end{document}